\def \PLB {{ Phys. Lett.} {\bf B}}
\def \NPB {{ Nucl. Phys.} {\bf B}}
\def \PL {{ Phys. Lett.}}
\def \CMP {{ Commun. Math. Phys. }}
\def \JMP {{ J. Math. Phys. }}
\def \NP {{ Nucl. Phys.}}

\def \IJMP {{ Int. J. Mod. Phys.}}

\def \bc {\begin{center}}
\def \ec {\end{center}}

\def \bfr {\begin{flushright}}
\def \efr {\end{flushright}}

\def \ba {\begin{array}}
\def \ea {\end{array}}

\def \bea {\begin{eqnarray}}
\def \eea {\end{eqnarray}}

\def \be {\begin{equation}}
\def \ee {\end{equation}}

\def \um {\frac{1}{2}}
\def \f {\frac}

\def \cL {{\cal L}}
\def\kbar{{\mathchar'26\mkern-9mu\lambda}}
\def \al {z}
\def \alc {\bar{z}}
\def \l[{\left[}
\def \r]{\right]}
\def \lt[{\left[}
\def \rt]{\right]}
\def \heta {h}

\def\w#1{\vec{v}_#1}
\def\z#1{\vec{z}_#1}
\def\u#1{\vec{u}_#1}

\def \W {{\cal W}}
\def \R {{\mathbb R}} 
\def \Z {{\mathbb Z}}
\def \N {{\mathbb N}}
\def \C {{\mathbb C}}
\def \d {d}

\def \nn {\nonumber}
\newcommand{\bra}[1]{ \left<#1\right| }
\newcommand{\ket}[1]{ \left|#1\right> }
\newcommand{\scprod}[2]{ \left<#1\right.|\left.#2\right> }

\documentclass[11pt,epsf]{article}
\usepackage{amssymb,amsfonts}
\usepackage{graphicx}
\begin{document}

\begin{center}
{\LARGE {\bf Higher-$U(2,2)$-spin fields and higher-dimensional
$\W$-gravities: quantum AdS space and radiation phenomena }}
\end{center}
\bigskip
\bigskip

\centerline{{\sc Manuel Calixto}}

\bigskip

\bc
{\it Departamento de Matem\'atica Aplicada y Estad\'\i stica,
Universidad Polit\'ecnica de Cartagena, Paseo Alfonso XIII 56, 30203 Cartagena, Spain}
\\ and \\
{\it Instituto de Astrof\'\i sica de Andaluc\'\i a,
Apartado Postal 3004, 18080 Granada, Spain}    \\

\bigskip

E-mail: Manuel.Calixto@upct.es
\ec

\bigskip

\bigskip
\begin{center}
{\bf Abstract}
\end{center}
\small

\begin{list}{}{\setlength{\leftmargin}{3pc}\setlength{\rightmargin}{3pc}}
\item A physical and geometrical interpretation of previously
introduced tensor operator algebras of $U(2,2)$ in terms of algebras of
higher-conformal-spin quantum fields on the anti-de Sitter space AdS$_5$
is provided. These are higher-dimensional ${\cal W}$-like algebras and
constitute a potential gauge guide principle towards the formulation
of induced conformal gravities (Wess-Zumino-Witten-like models) in
realistic dimensions. Some remarks on quantum (Moyal) deformations
are given and potentially tractable versions of noncommutative AdS spaces are
also sketched. The role of conformal symmetry in the microscopic
description of Unruh's and Hawking's radiation effects is discussed.
\end{list}
\normalsize



\newpage

\section{Introduction}

A consistent and feasible quantum theory of the gravitational
field is still lacking, but we know of some crucial tests that such
a candidate theory ought to pass.  Mainly, it should account
for radiation phenomena,  like black hole evaporation,
and it should exhibit a critical behaviour at
short distances (Planck scale), where the intrinsic structure of space-time and
the physics we know should radically change. Actually, both the Bekenstein-Hawking formula
for black hole entropy $S=\frac{k_B\Sigma}{4G\hbar}$ ---and temperature
(\ref{tempHawking})--- and Planck length $\kbar=\sqrt{\hbar G/c^3}$
provide a remarkable mixture of: quantum mechanics (Planck's constant
$\hbar$), gravity (Newton's constant $G$), relativity (velocity of light
$c$), thermodynamics (Boltzmann's constant $k_B$), and geometry (the
horizon area $\Sigma$).

A statistical mechanical explanation of black hole thermodynamics in terms
of counting of microscopic states has been recently
given in \cite{Carlipcqg}. According to this reference,
there is strong evidence that conformal
field theories provide a universal (independent of details of the particular
quantum gravity model) description of low-energy black hole entropy, which
is only fixed by symmetry arguments. The Virasoro algebra turns out to be the
relevant subalgebra of surface deformations of the horizon of an arbitrary
black hole and constitutes the general
gauge (diffeomorphism) principle that governs the density of states.

Although surface deformations appear as a constraint algebra, under which one
might expect all the physical states on the horizon to be singlets, quantum
anomalies and boundary conditions introduce central charges and change this
picture, thus causing gauge/diffeomorphism modes to become physical along the
horizon.\footnote{This seems to be an important and general feature of quantum
gauge theories as opposite to their classical counterparts. For example,
see Refs. \cite{ymas,gtp,accgmgt} for a cohomological ({\it Higgs-less}) generation
of mass in Yang-Mills theories through non-trivial representations of the gauge
group; in this proposal, gauge modes become also physical and the corresponding
extra field degrees of freedom are transferred to the vector potentials (longitudinal
components) to form massive vector bosons.}
In this way, the calculation of thermodynamical quantities, linked to the statistical
mechanical problem of counting microscopic states,  is reduced
to the study of the representation theory and central charges
of a relevant symmetry algebra.

Unruh effect \cite{Unruh} (vacuum radiation in uniformly accelerated frames)
is another interesting physical phenomenon linked to the previous one. A statistical
mechanical description (from first principles) of it has also been given in
\cite{conforme} (see also Section \ref{2ndgaqsec3} below) and related to the
dynamical breakdown of part of the conformal symmetry: the special conformal
transformations (usually interpreted as transitions to a uniformly relativistic
accelerated frame), in the context of conformally invariant quantum field theory.
The Unruh effect can be considered as a ``first order effect'' that gravity has
on quantum field theory, in the sense that transitions to
uniformly accelerated frames are just enough to account for it. To account for
higher-order effects one should consider more general diffeomorphism algebras, but
the infinite-dimensional character of conformal symmetry seems to be an exclusive
patrimony of two-dimensional physics, where the Virasoro algebra provides
the main gauge guide principle. This statement is not rigorously true, and the
present article is intended to provide higher-dimensional analogies of
the infinite two-dimensional conformal symmetry, which could be useful as potential
gauge guiding principles towards the formulation of gravity models in realistic
dimensions.

Actually, the so called $w$-algebras constitute a generalization
of the Virasoro symmetry, and two-dimensional $w$-gravity models, generalizing
Polyakov's induced gravity, have been formulated (see next Section). They turn out
to be constrained Wess-Zumino-Witten models. The algebras
$w$ have also a space origin as (area preserving) diffeomorphisms and Poisson
algebras of functions on symplectic manifolds (e.g. cylinder).
There is a group-theoretic structure underlying their quantum (Moyal \cite{Moyal}) deformations
(collectively denoted by $\W$), according to which $\W$ algebras are just
particular members of a one-parameter family $\cL_\rho(sl(2,\mathbb{R}))$
---in the notation of the present paper---
of non-isomorphic infinite-dimensional Lie-algebras of
$SL(2,\mathbb{R})$ tensor operators ---see later on Eq. 
(\ref{sl2rtensorop}). The connection with the theory of
higher-spin gauge fields in (1+1)- and (2+1)-dimensional anti-de Sitter space
AdS \cite{Fradkin} ---homogeneous spaces of $SO(1,2)\sim SL(2,\mathbb{R})$ and $SO(2,2)\sim
SL(2,\mathbb{R})\times SL(2,\mathbb{R})$, respectively--- is then apparent in this
group-theoretical context. The AdS spaces are arousing an increasing interest
as asymptotic background spaces in (super)gravity theories, essentially sparked off by
Maldacena's conjecture \cite{Maldacena}, which establishes a correspondence
of holographic type between field theories on AdS and conformal
field theories on the boundary (locally Minkowski). The AdS space plays also
an important role in the above mentioned attempts to understand the
microscopic source of black hole entropy.

This scenario constitutes a suitable approach for our purposes. Indeed,
the (3+1)-dimensional generalization of the previous constructions is
just straight-forward when considering the infinite-dimensional Lie algebras
$\cL_{\vec{\rho}}(so(4,2))$ of $SO(4,2)$-tensor operators (where $\vec{\rho}$ is now a
three-dimensional vector). They can be regarded as
infinite enlargements of the (finite) conformal symmetry $SO(4,2)$ in 3+1 dimensions
which incorporate the subalgebra ${\rm diff}(4)$ of diffeomorphisms
[the higher-dimensional analogue of the Virasoro algebra diff$(S^1)$]
of the four-dimensional
space-time manifold (locally Minkowski), in addition to interacting fields with
all  $SO(4,2)$-spins. This fact makes $\cL_{\vec{\rho}}(so(4,2))$ a potential
gauge guide and a main ingredient towards the formulation of
higher-dimensional (Wess-Zumino-Witten-like) gravity models.

The classification and labelling of tensor operators of Lie groups
other than $SL(2,\mathbb{R})$ and $SU(2)$ is not an easy task in
general. In Refs. \cite{infdimal,v3p1}, the author provides a
quite appropriate basis of tensor operators for
$\cL_{\vec{\rho}}(u(N_+,N_-)),
\vec{\rho}=(\rho_1,\dots,\rho_{N})$, $N \equiv N_++N_-$,  and
calculates the structure constants for the particular case of the
boson realization of quantum associative operator algebras on
algebraic (K\"ahler) manifolds
$F_{{}_{N_+N_-}}=U(N_+,N_-)/U(1)^{N}$, also called {\it flag}
manifolds (see Refs. \cite{Helgason,Pressley} for a formal
definition of flag manifolds). We are just interested in the
particular case of $U(2,2)\simeq (SO(4,2)\times
U(1))/\mathbb{Z}_4$. Tensor labelling coincides here with the
standard pattern for wave-functions in the carrier space of
unirreps of $U(N)$ (see Appendix).

In this article we look at this abstract, purely algebraic,
construction from a more physical, field-theoretic, point of view.
In particular, we shall show how to recover $\cL_{\vec{\rho}}({\cal G})$ from
quantum field theory on homogeneous spaces $M=G/P$, formulated as a second
quantization on  the group $G$. The particular cases of $G=SO(1,2)\sim
SU(1,1)$ and $G=SO(4,2)\sim SU(2,2)$ will give us the (generalized)
$w$-structures as algebras of
higher-conformal-spin quantum fields on anti-de Sitter spaces.
One can interpret $\cL_{\vec{\rho}}({\cal G})$ as the quantum analogue of the
Poisson and symplectic (volume-preserving) diffeomorphism algebras,
$C^\infty(F)$ and ${\rm sdiff}(F)$, of functions $\psi$
(higher-$G$-spin fields) on coadjoint orbits $F$ of $G$, with a given associative
and non-commutative $\star$-product. The Planck length $\kbar$ enters here as
a deformation parameter and it could be basically interpreted as
an upper-resolution bound to any position measurement in any quantum
gravity model. The ideas of
Non-Commutative Geometry (NCG) apply perfectly in this picture,
providing ``granular''
descriptions of the underlying `quantum' space: the non-commutative analogue of $M$
(see \cite{Madore} for similar concepts). The
classical (commutative) case is recovered in the limit $\kbar\to 0$ (large scales) and
$\rho\to\infty$ (high density of points), so that `volume elements' remain finite.
In this limit, the classical geometry is recovered as:
\be
\lim_{\kbar\to 0,\rho\to\infty} {\cal L}_\rho({\cal G})\simeq
C^\infty(F)\subset
{\rm sdiff}(F), \label{classlim}
\ee
where $F$ is the (generalized) phase-space associated with the space-time $M$.
We shall discuss  all these structures in more detail in Sec. \ref{wbrief} and Sec. \ref{qadse}.

The organization of the paper is as follows. The next Section is devoted to a
brief excursion through $w$ algebras, their quantum deformations $\W$ and their underlying
group-theoretic nature. Moyal $\star$-products are discussed in Sec. \ref{nctorus},
in connection with the particular case of the non-commutative torus.
In Sec. \ref{2ndgaqsec}, a
general approach to the quantization of fields in homogeneous spaces
$M=G/P$ from a second quantization on a group $G$ is exposed. The simple
cases of de Sitter dS$_2\simeq SO(2,1)/SO(1,1)$ and anti-de Sitter
AdS$_2 \simeq SO(1,2)/SO(1,1)$ spaces in 1+1 dimensions are explicitly
laid out in Sec. \ref{2ndgaqsec2}, leaving the (4+1)-dimensional
case AdS$_5\simeq SO(4,2)/SO(4,1)$ as an Appendix, where
the unitary irreducible representations of $U(2,2)$ are explicitly
given in terms of holomorphic wave functions on flag (K\"ahler) manifolds
$F_{2,2}=U(2,2)/U(1)^4$. The interesting
physical effects of vacuum radiation phenomena
are discussed in Sec. \ref{2ndgaqsec3} in the context of a quantum
field theory on AdS$_2$, representing the left-(or right-)moving sector
of a conformal field theory in 1+1 dimensions; the spectrum of outgoing
particles is exactly calculated and is proven to be a generalization of the
black body (Planckian) spectrum, this recovered as a given limit.
The embedding of quantum field operators and space-time symmetries in
a larger structure (higher-dimensional $\W$ algebras) containing
the diffeomorphisms of the space-time manifold (higher-dimensional analogue of
the Virasoro algebra) and all fields with
arbitrary (generalized) spin is given in Sec. \ref{v3p1}. A geometrical
interpretation in terms of Poisson (and symplectic diffeomorphism) algebras
$C^\infty(F)$ of functions $\psi$ on
algebraic manifolds $F$ (coadjoint orbits) of a group $G$
and its quantum deformations $\cL_{\vec{\rho}}(G)$ is also
discussed. These quantum deformations can be seen as non-commutative $C^\star$-algebras with
an associative and non-commutative $\star$-product.
This prepares us to define what we mean by quantum (non-commutative)
AdS space in this context. The last
Section is devoted to conclusions and outlook.

\section{$\W$ algebras and two-dimensional quantum space \label{wbrief}}

In the last decade, a large body of literature has been devoted to the study
of the so-called $\W$-algebras. These algebras were first introduced as
higher-spin extensions of the Virasoro algebra \cite{Zamolodchikov}
through the operator product expansion of the stress-energy
tensor and primary fields in two-dimensional conformal field theory.
$\W$-algebras have been widely used in two-dimensional physics, mainly in
condensed matter (quantum Hall effect), integrable models
(Korteweg-de Vries, Toda), phase transitions in two dimensions,
stringy black holes and, at a more fundamental level, as the underlying
gauge symmetry  of two-dimensional gravity models
generalizing the Virasoro gauged symmetry in the light-cone discovered
by Polyakov \cite{Poly2,Poly} by adding spin $>2$ currents (see e.g.
\cite{Bergshoeff} and \cite{Shen,Hull} for a review).

Only when all (infinite) conformal spins $s\geq 2$ are considered,
the algebra (denoted by $w_{\infty}$) is proven to be of Lie type;
moreover, currents of spin $s=1$ can also be included, thus
leading to the Lie algebra $w_{1+\infty}$, which plays a
determining role in the classification of all universality classes
of incompressible quantum fluids and the identification of the
quantum numbers of the excitations in the quantum Hall effect
\cite{Capelli}.

As already said, the algebras $w$ prove to have a space-time
origin as (symplectic) diffeomorphism algebras and Poisson
algebras of functions $\psi$ on symplectic manifolds. For example,
$w_{1+\infty}$ is related to the algebra of diffeomorphisms of the
cylinder. In fact, let us choose the next set of classical
functions of the bosonic (harmonic oscillator) variables
$a=\frac{1}{\sqrt{2}}(q+ip),\,\bar{a}=\frac{1}{\sqrt{2}}(q-ip)$
(we are using mass and frequency $m=1=\omega$, for simplicity):
\be
L^I_{|n|}\equiv\um a^{2|n|}(a\bar{a})^{I-|n|},\,\,\,
L^J_{-|m|}\equiv\um \bar{a}^{2|m|}(a\bar{a})^{J-|m|},\label{auaral}
\ee
where $n,m\in \mathbb Z/2; I,J\in \mathbb Z^+/2$.
A straightforward calculation from the
basic Poisson bracket $\{a,\bar{a}\}=i$ provides the following formal
Poisson algebra:
\be
\{L^I_m,L^J_n\}=-i[Jm-In]L^{I+J-1}_{m+n}\,,
\label{auaralcom}
\ee
of functions $L$ on a two-dimensional phase space (see \cite{simplin,WB}).
The (conformal-spin-2)
generators $L_n\equiv L^1_n$ close the
Virasoro algebra without central extension,
\be
\{L_m,L_n\}=i(n-m)L_{m+n},
\ee
and the (conformal-spin-1) generators
$\phi_m\equiv L^{0}_m$ close the non-extended Abelian Kac-Moody
algebra of ``string-modes'',
\be
\{\phi_m, \phi_n\}=0.
\ee
In general, the higher-spin fields $L^{I}_n$ have
conformal-spin $s=I+1$ and conformal-dimension $n$ (the eigenvalue of
$L^1_0$).

Induced actions for these ``$w$-gravities'' have been written (see
for example \cite{Bergshoeff}), which turn out to be constrained
Wess-Zumino-Witten models \cite{Nissimov}, as happens with standard induced
gravity. The quantization
procedure {\it deforms} the classical algebra $w$ to the quantum
algebra $\W$ due to the presence of anomalies ---deformations of
Moyal type of Poisson and symplectic-diffeomorphism algebras caused
essentially by normal order ambiguities (see Sec. \ref{nctorus}).
Also, generalizing the
$SL(2,\R)$ Kac-Moody hidden symmetry of Polyakov's induced
gravity, there are $SL(\infty,\R)$ and  $GL(\infty,\R)$
Kac-Moody hidden symmetries for $\W_{\infty}$ and  $\W_{1+\infty}$ gravities,
respectively \cite{Popehidden}.

The group-theoretic structure underlying these $\W$
algebras was elucidated in \cite{Pope}, where  $\W_{\infty}$ and
$\W_{1+\infty}$ appeared to be distinct members ($\rho=0$ and
$\rho=-1/4$ cases, respectively) of a one-parameter
family $\cL_\rho(sl(2,\R))$
of non-isomorphic \cite{Hoppe2} infinite-dimensional Lie-algebras
of $SL(2,\R)\simeq SU(1,1)/\Z_2$ tensor
operators
\be
\hat{L}^I_{\pm |m|}\sim \underbrace{\l[ \hat{L}_{\mp},\l[ \hat{L}_{\mp},\dots
\l[ \hat{L}_{\mp},\right.\right.\right.}_{I-|m|\,\,{\rm times}}
\left.\left.\left.
(\hat{L}_\pm)^I\r] \dots\r] \r] =
({\rm ad}_{\hat{L}_{\mp}})^{I-|m|}(\hat{L}_\pm)^I,\label{sl2rtensorop}
\ee
where the $su(1,1)$ Lie-algebra generators $\hat{L}_{\pm}, \hat{L}_{0}$
fulfil the standard commutation relations
\be
\l[\hat{L}_\pm,\hat{L}_0\r]=\pm
\hbar\hat{L}_\pm\,,\;\;\;\;\;
\l[\hat{L}_+,\hat{L}_-\r]=2\hbar\hat{L}_0.
\ee
In  more formal language, $\cL_\rho(su(1,1))$ is the factor algebra
$\cL_\rho(su(1,1))={\cal U}(su(1,1))/{\cal I}_\rho$
of the universal enveloping
algebra ${\cal U}(su(1,1))$ by the ideal
${\cal I}_\rho=(\hat{C}-\hbar^2\rho){\cal U}(su(1,1))$
generated by the Casimir operator
$\hat{C}=(\hat{L}_0)^2-\um(\hat{L}_+\hat{L}_-+\hat{L}_-
\hat{L}_+)$ of $su(1,1)$
($\rho$ denotes an arbitrary complex number). This simply means that
we substitute the Casimir operator $\hat{C}$ by the constant $\hbar^2\rho$ whenever
it appears in the commutator (structure constants).
The structure constants for $\cL_\rho(su(2))$ and
$\cL_\rho(su(1,1))$ are well known for the Racah-Wigner basis of
tensor operators \cite{Biedenharn}, and they can be written in terms of
Clebsch-Gordan and (generalized) $6j$-symbols \cite{Hoppe,Pope,Fradkin2}.
Another interesting feature of $\cL_\rho(su(2))$ ---or its non-compact
version $SU(1,1)$--- is that, when $\rho$
coincides with the eigenvalue of $\hat{C}$ in an irrep $D_j$ of
$SU(2)$, that is $\rho=j(j+1)$, there exists and ideal $\chi$ in
$\cL_{\rho}(su(2))$ such that
the quotient $\cL_{\rho}(su(2))/\chi\simeq sl(2j+1,C)$ or $su(2j+1)$ ---by
taking a compact real form of the complex Lie algebra \cite{Burnside}.
That is, for $\rho=j(j+1)$ the infinite-dimensional algebra
$\cL_{\rho}(su(2))$ {\it collapses} to a finite-dimensional one (a matrix algebra).
This fact was used in \cite{Hoppe} to approximate
$\lim_{\stackrel{\rho\to\infty}{\hbar\to 0}}\cL_\rho(su(2))
\simeq {\rm sdiff}(S^2)$
by $su(N)|_{N\to\infty}$ (``large number of colours'') in the relativistic spherical
membrane. A physical
interpretation of this ``collapse phenomenon'' could be given in the
context of Non-Commutative Geometry, which provides a finite, ``fuzzy or
cellular'' description of the (non-commutative) space \cite{Madore,Connes}.
A simple example  will be discussed in Section \ref{nctorus}
in connection with the {\it noncommutative torus} \cite{nctorusref}.

Noncommutative geometry is said to be a `pointless' geometry,
where the notion of a pure state $\psi$ (more precisely, its `covariant symbol' $\hat{\psi}$)
on the manifold $M=G/P$ replaces that of a point.
In this sense, the parameter $\rho$ above could be conceived as
a `density of (quantum) points' (e.g., the dimension of the corresponding representation).
Actually, just as the standard differential geometry of $M=S^2$ can be described by using the
(commutative) algebra $C^\infty(M)$ of smooth complex functions $\psi$ on $M$ (spherical
harmonics), a noncommutative
geometry for $M$ can take $\cL_\rho({\cal G})$ (seen as an associative
algebra with a noncommutative $\star$-product) as the starting point. For the
critical values $\rho=d$ (the dimension of an irrep of $G=SU(2)$), the `quantum sphere'
has a finite number of quantum points per unit volume, as the infinite-dimensional
algebra $\cL_\rho({\cal G})$ collapses to a (finite) complex matrix algebra.
The classical geometry on $M$ is then recovered in the particular
limit (\ref{classlim}) ---i.e. high density of quantum points and small quantum cells:
the Planck area $\kbar^2$--- upon which, the commutator $[\cdot,\cdot]$
becomes the usual Poisson bracket $\{\cdot,\cdot\}$ on the sphere.
It is worth-mentioning that the whole idea was noticed
long time ago by Dirac \cite{Dirac}, who realized the
possibility of describing phase-space
physics in terms of the quantum analogue of the algebra of functions (the covariant symbols)
and the absence of localization expressed by the Heisenberg uncertainty principle.
As happens in phase-space, the fundamental scale (Planck length) $\kbar$
should establish an upper-resolution bound to any position measurement in any quantum
gravity model.

Let us illustrate briefly this ``collapse" phenomenon with the simple
example of (a subalgebra of) the $\W$ algebra $\cL_\rho(su(1,1))$ in connection with 
the noncommutative torus.

\subsection{Noncommutative Torus\label{nctorus}}

The noncommutative torus \cite{nctorusref} is one of the basic examples of a noncommutative
geometry \cite{Madore,Connes,Landi}  which captures features of the difference between an ordinary manifold
and a `noncommutative space'. Recent interest in such geometries has occurred in the
physics literature in the context of their relation to M-Theory 
\cite{ConnesMT}.

Let us consider the following new set of classical functions of the bosonic variables
$a(\bar{a})=\frac{2\pi}{\ell}(x_1\pm ix_2)$:
\be {L}_{\vec{n}}\equiv e^{\frac{2\pi i}{\ell} \vec{n}\cdot\vec{x}} =
\sum_{I=0}^\infty\sum_{l=0}^I 2(-1)^{I}
\frac{(n_1+in_2)^{I+l}}{2^{I+l}(I+l)!}\frac{(n_1-in_2)^{I-l}}{2^{I-l}(I-l)!}
\,\,L^I_l, \label{torusfunc}
\ee
obtained from (\ref{auaral}), where $\vec{x}=(x_1,x_2)$
is a pair of real coordinates (modulo $\ell$)
and $\vec{n}=(n_1,n_2)$ is a pair of integer numbers.
We identify (\ref{torusfunc}) with the set $C^\infty(T^2)$ of
smooth functions on a two-dimensional torus,
which is embedded in the set of functions (\ref{auaral})
on a cylinder [note that the lower-index $l$ of $L^I_l$ in (\ref{torusfunc}) is
restricted by $0\leq l\leq I$, whereas it can take any (half-)integer
value in (\ref{auaral})]. We shall restrict ourselves to this
subset of the whole $w$-algebra  (\ref{auaralcom})
where the above-mentioned ``collapse" phenomenon will be more apparent.

The ordinary product of functions $L_{\vec{m}}\cdot L_{\vec{n}}=L_{\vec{m}+\vec{n}}$ defines
$C^\infty(T^2)$ as a commutative algebra. We can assign a (Hamiltonian) vector field
$H_{\vec{m}}\equiv \{ L_{\vec{m}},\cdot\}_{{\rm P}}$ to any function $L_{\vec{m}}$, where:
\be
\left\{  L_{\vec{m}}, L_{\vec{n}}\right\}_{{\rm P}}= P^1(
L_{\vec{m}}, L_{\vec{n}})= \Upsilon_{jk}\frac{\partial
L_{\vec{m}}}{\partial x_j} \frac{\partial  L_{\vec{n}}}{\partial
x_k}= \frac{4\pi^2}{\ell^2}\vec{n}\times\vec{m}\,
L_{\vec{m}+\vec{n}}\,, \label{Poisson} \ee denotes the Poisson
bracket and  $\Upsilon_{2\times 2}\equiv \left(\begin{array}{cc} 0
& 1 \\ -1 &0\ea\right)$ is the symplectic form on the torus. The
vector fields $H_{\vec{m}}$ constitute a basis of symplectic
diffeomorphisms sdiff$(T^2)$, that is, they preserve the area
element $dx_1\wedge dx_2$ of the torus.

The quantum analogue of $C^\infty(T^2)$ can be captured from a classical construction by extending the
Poisson bracket (\ref{Poisson}) to its deformed version: the Moyal bracket,
\be
\left\{ L_{\vec{m}}, L_{\vec{n}}\right\}_{{\rm M}}=
 L_{\vec{m}}\star L_{\vec{n}}- L_{\vec{n}}\star L_{\vec{m}}
=\sum_{r=0}^{\infty}
\frac{2}{(2r+1)!}\left(\frac{\kbar^2}{2\pi i}\right)^{2r+1}
P^{2r+1}( L_{\vec{m}}, L_{\vec{n}}),\label{Moyaleq}
\ee
where
\be
 L_{\vec{m}}\star L_{\vec{n}}
\equiv\exp(\frac{\kbar^2}{2\pi i} P)( L_{\vec{m}}, L_{\vec{n}})=
e^{2\pi i\frac{\kbar^2}{\ell^2}\vec{m}\times\vec{n}}  L_{\vec{m}+\vec{n}}
\label{star}
\ee
is an invariant, associative and noncommutative  $\star$-product, and
by $P^r$ we mean
\bea
P^r(L,L')&\equiv&\Upsilon_{j_1k_1}\dots\Upsilon_{j_rk_r}
\frac{\partial^r L}{\partial x_{j_1}\dots\partial
x_{j_r}}\frac{\partial^r L'}{\partial x_{k_1}\dots\partial
x_{k_r}}\nn\\
&=& \sum_{l=0}^r(-1)^l\left(\ba{c} r\\ l\ea\right)
[\partial^{r-l}_{x_1}\partial^{l}_{x_2} L]
[\partial^{r-l}_{x_2}\partial^{l}_{x_1} L'] \label{star2}
\eea
where $P^0(L,L')\equiv L\cdot L'$ denotes the ordinary (commutative) product of functions.

The $\star$-product (\ref{star}) also admits an integral representation:
\be
(L\star L')(\vec{x})=\frac{i\ell^2}{4\pi\kbar^6}\int_0^\ell
d\vec{x}'d\vec{x}''e^{-\frac{2\pi i}{\kbar^2}|\vec{x}\vec{x}'\vec{x}''|}
L(\vec{x}')L'(\vec{x}'') \label{convotorus}
\ee
with
\be
|\vec{x}\vec{x}'\vec{x}''|\equiv \vec{x}\times\vec{x}'+
\vec{x}'\times\vec{x}''+\vec{x}''\times\vec{x}\,,
\ee
which is interesting when one wants to extend the
$\star$-product to other symplectic manifolds like coadjoint orbits
of certain groups (see Sec. \ref{qadse}).

The Moyal bracket (\ref{Moyaleq}) reminds us a commutator between operators
\be
\left\{L_{\vec{m}},L_{\vec{n}}\right\}_{\rm M}=
\left[\hat{L}_{\vec{m}},\hat{L}_{\vec{n}}\right]\equiv
\hat{L}_{\vec{m}}\star\hat{L}_{\vec{n}}-\hat{L}_{\vec{n}}
\star\hat{L}_{\vec{m}}=
2i\sin\left(2\pi \frac{\kbar^2}{\ell^2} \vec{m}\times\vec{n}\right)
\hat{L}_{\vec{m}+\vec{n}} \,.\label{Moyalcom}
\ee
Thus, this $\star$-product equips the set (\ref{torusfunc}) with a
noncommutative $C^\star$-algebra structure,
which we shall denote by $C^\star_\rho(T^2)$.
Just as the standard geometry of $T^2$ can be described by using the
algebra $C^\infty(T^2)$ of smooth complex functions (\ref{torusfunc}) on
$T^2$ with the ordinary (commutative) product, a noncommutative
geometry for $T^2$ can be described
by using its ``quantum" analogue $C^\star_\rho(T^2)$.
Noncommutative geometry offers a broader spectrum of
possibilities. In fact,
let us see how the definition of ``quantum torus" is
richer than that of the (standard) ``classical torus",
which eventually comes up as a particular
limiting case of the former one.

Note that, when the surface of the torus $\ell^2$ contains an integer
number $q$ of times the {\it minimal cell} $\kbar^2$ (that is,
$\ell^2=q \kbar^2$), the infinite-dimensional algebra
(\ref{Moyalcom}) collapses to a finite-dimensional matrix algebra: the Lie
algebra of the unitary group $U(q/2)$ for $q$ even or
$SU(q)\times U(1)$ for $q$ odd (see \cite{Fairlie}).
In fact, taking the quotient in (\ref{Moyalcom}) by the equivalence
relation  $\hat{L}_{\vec{m}+q\vec{a}}\sim
\hat{L}_{\vec{m}},\,\forall \vec{a}\in {\mathbb Z}\times \mathbb Z$, it can be seen that
the following identification $\hat{L}_{\vec{m}}=\sum_{k}
e^{\frac{2\pi i}{q}m_1k}X_{k,k+m_2}$
implies a change of basis in the step-operator Lie-algebra (\ref{pun})
of $U(q)$.

Thinking of $\rho=\frac{\ell^2}{\kbar^2}$ as a `density of quantum points',
we can conclude that: for the critical values $\rho_c=q\in \mathbb Z$, the Lie algebra
(\ref{Moyalcom}) is {\it finite};
that is, the quantum analogue of the torus has a `finite number $q$ of
quantum points'. It is in this sense that we talk about a
`cellular structure of space'. Actually, given the formal basic commutator
$\left[ x_1,x_2\right]=-i\kbar^2/\pi$ between ``position operators" on the torus,
this cellular structure is a
consequence of the absence of localization expressed by the Heisenberg
uncertainty relation $\Delta x_1 \Delta x_2 \geq \kbar^2/(2\pi)$
(see Figure \ref{nctorusfig}).

\begin{figure}[htb]
\begin{center}
\includegraphics[height=5cm,width=10cm]{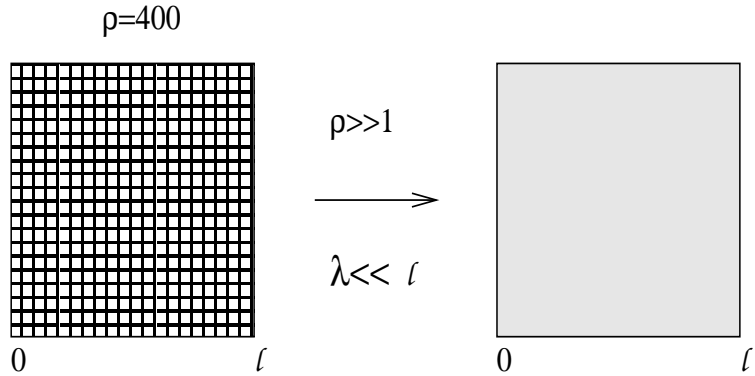}
\caption{{ {\sf In extreme quantum conditions $\ell\sim\kbar$ (size of the space comparable to
the Planck length), the cellular structure of space could be conspicuous.}}}
\label{nctorusfig}
\end{center}
\end{figure}

In the (classical) limit of large number of points $\rho\to\infty$ and
$\kbar\to 0$ (such that the size $\ell^2=\rho\kbar^2$ of the torus remains finite) we
recover the original (commutative) geometry on the torus.
For example, it is easy to see that
\be
\lim_{\rho\to \infty,\,\kbar\to 0}
\frac{i\pi}{\kbar^2}\left[\hat{L}_{\vec{m}},\hat{L}_{\vec{n}}\right]
=\frac{4\pi^2}{\ell^2}\vec{n}\times\vec{m}\,{L}_{\vec{m}+\vec{n}}=
P^1({L}_{\vec{m}},{L}_{\vec{n}})
\ee
coincides with the (classical) Poisson bracket (\ref{Poisson}) of functions in
$C^\infty(T^2)$. In particular, the last equality states that the Poisson
algebra (\ref{Poisson}) formally coincides with the Lie algebra of the
group of infinite unitary matrices $U(\infty)$ \cite{Fairlie}. This is just
a facet of the general problem of approximating
infinite-dimensional Lie algebras of symplectic diffeomorphisms on
homogeneous manifolds by large-$N$ matrix algebras.

This simple example gives us a taste of what should happen in higher
dimensions and less trivial quantum manifolds. Although the general subject of
Noncommutative Geometry is rather developed (see e.g.
\cite{Connes,Landi,Madore}), there is still a technical difficulty partially due
to the lack of tractable (yet non-trivial) noncommutative versions of curved
spaces. Let us proceed, firstly, by showing how to obtain quantum fields on a
homogeneous space $M=G/P$ and, secondly, how to embed those fields in a
noncommutative structure (``higher-dimensional $\W$-algebra")
living in the universal enveloping algebra of $G$.

\section{QFT in Homogeneous Spaces from a Second
Quantization on a Group\label{2ndgaqsec}}

The sequel to the previous constructions will consist in
generalizing the concept of $\W$-algebra to realistic dimensions.
Once the group-theoretic nature of $\W$-algebras is elucidated,
there is no conceptual difficulty in jumping from 1+1 to 3+1
dimensions. Indeed, we should just substitute $G=SU(1,1)$ by
$G=SU(2,2)$ (the conformal group $SO(4,2)$ in 3+1 dimensions,
except for discrete symmetries) and consider now
$\cL_{\vec{\rho}}(su(2,2))$ as a higher-dimensional generalization
of the $\W$-algebras. However, there is a technical difficulty;
the generalization of these constructions to general unitary
groups proves to be quite unwieldy, and a canonical classification
of $U(N)$-tensor operators has, so far, been proven to exist only
for $U(2)$ and $U(3)$ (see \cite{Biedenharn} and references
therein). As already mentioned, tensor labelling is provided in
these cases by the Gel'fand-Weyl pattern for vectors in the
carrier space of the irreps of $U(N)$. A step forward in this
direction is given in the letter \cite{infdimal}, where the author
provides a quite appropriate basis of operators for
$\cL_{\vec{\rho}}(u(N_+,N_-)),
\vec{\rho}=(\rho_1,\dots,\rho_{N})$, $N \equiv N_++N_-$,  and
calculates the structure constants, for the particular case of the
boson realization of quantum associative operator algebras on
algebraic manifolds $F_{{}_{N_+N_-}}=U(N_+,N_-)/U(1)^{N}$ ({\it
flag} manifolds). This is a particular, yet important, case inside
the $N$-parameter family $\cL_{\vec{\rho}}(u(N_+,N_-))$, where the
standard Moyal bracket \cite{Moyal} provides the primary quantum
deformations (i.e. an invariant associative $\star$-product). The
general case is still under study, although Moyal deformation
still captures much of the essence of the full quantum algebra, as
happens to deformations of $\W$ algebras \cite{Fairlie2}.

The aim of this article will not be to provide an exhaustive study of quantum deformations
$\cL_{\vec{\rho}}(u(2,2))$ but, rather, to give a physical and geometrical interpretation
of the classical limit $\cL_{\infty}(G)$ for $G=U(1,1)$ and $G=U(2,2)$ in terms of
higher-$G$-spin algebras of fields $\psi$ on anti-de Sitter spaces
in 1+1 and 4+1 dimensions, respectively. For this, let us outline how to formulate
a quantum field theory in a homogeneous space $M=G/P$ as a second quantization on a
group $G$. Some comments on quantum deformations will be given in Section \ref{qadse}.

\subsection{Second Quantization on a Group\label{2ndgaqsec1}}

In the standard formulation, the Quantum Mechanics of a particle
in a curved space $M$ is explicitly constructed in the configuration
space (space image), making use of the intrinsic differentiable structure
of $M$ (generally, a pseudo-Riemann, globally hyperbolic manifold with
metric $g^{\mu\nu}$). However, when $M$ is symmetrically enough or, more
precisely, when $M$ is somehow embedded in a symmetry group $G$, this
case offers more possibilities (see e.g. \cite{coadjoint1,Woodhouse,Berezin,GAQ,Isham,infiel} for classical references 
on group-theoretical methods in canonical and geometric quantization). In fact, we can always use the large
amount of geometrical and algebraic resources that a group offer to us:
left- and right-invariant vector fields, Casimir operators, invariant forms
(e.g. volume elements), symplectic structures on the coadjoint orbits, results
of the representation theory, etc; in particular,
the problem of calculating the
general solution to the equations of motion, like that of a free scalar 
field in a curved background,
\be
(g^{\mu\nu}\nabla_\mu\nabla_\nu+m^2+\xi R)\phi(x)=0\label{fieldeq}
\ee
($\nabla_\mu$ denotes the covariant derivative), 
is reduced to finding the unitary and irreducible representations with a
fixed value of the Casimir operator, 
$\hat{C}\sim \frac{\partial^2}{\partial t^2}+K$,  on a given support space
of constant curvature given by the eigenvalue ($k$) of $\hat{C}$ (for example,
$\hat{C}=\partial_\mu\partial^\mu$ for the Poincar\'e group). \footnote{The method applies to arbitrary 
spin fields, as different representations of $G$ with different eigenvalues $s(s+1)$ of the 
Casimir operator of the rotation (sub)group $SU(2)$. See the Appendix.}
Such is the case of de Sitter and  anti-de Sitter spaces in
$d+1$ dimensions:
\be
{\rm dS}_{1+d}\simeq SO(d+1,1)/SO(d,1)\,,\;\;\; {\rm AdS}_{1+d}\simeq
SO(d,2)/SO(d,1)\,,\label{AdS}
\ee
which (mainly AdS) are arousing an increasing interest as asymptotic
background spaces in (super)gravity theories, as already said in the Introduction.
Let us outline
a general method for constructing a quantum field theory in a homogeneous
space $M$ of a group $G$.

Let ${\cal H}_s(G)=\{\psi^{(s)}_k\}_{k\in I}$  (the lower-index $k$ represents
quantum numbers that lie on a set $I$ for a given $G$-spin label $s$)
be an irreducible representation of $G$ obtained
from the regular representation, that is, the space of arbitrary complex functions
$\psi:G\to\mathbb{C},\, \psi(g)\in \mathbb{C},\,g\in G$).
Given the left action of $G$ on itself $L_{g'}(g)=g'\bullet g$
(where  $\bullet$ denotes the composition law),
the finite action $\rho$ of $G$ on ${\cal H}_s(G)$
\be
\psi'(g)\equiv (\rho(g')\psi)(g)=\psi(L_{g'}^{-1}(g))=
\psi(g'^{-1}\bullet g)\label{repre}
\ee
defines a unitary representation with respect to the scalar product:
\be
\langle \psi |\psi'\rangle\equiv \int_G{{d}^Lg \,
\bar{\psi}(g)\psi'(g)}\,,\label{scalarprod}
\ee
where ${d}^Lg=\theta^{L1}\wedge\overbrace{\dots}^{r={\rm dim}(G)}\wedge
\theta^{Lr}$ denotes the left-invariant integration measure,
which is constructed as an exterior product of left-invariant one-forms
$\theta^{L\beta}=\Lambda^\beta_\alpha(g) d g^\alpha$, and
$\{g^\alpha\}_{\alpha=1}^r$ is a system of coordinates on $G$.
The matrices  $\Lambda^\beta_\alpha(g)$ are calculated by duality
$\theta^{L\beta} (X^L_\alpha)=\delta^\beta_\alpha$ with the
left-invariant vector fields
\be
X^L_\alpha\equiv L_\alpha^\beta(g)
\frac{\partial}{\partial g^\beta}\,,\;\;\; L_\alpha^\beta(g)=
\left.\frac{\partial (g\bullet g')^\beta}{\partial
g'^\alpha}\right|_{g'=e}\,,\label{izquierdo}
\ee
where $e$ denotes the identity element; that is, $\Lambda^\beta_\alpha=
(L^{-1})^\beta_\alpha$. In general, the finiteness of this measure is not
guaranteed; however, this definition will be enough for our purposes.

For simplicity, we shall restrict ourselves to real quantum fields
\be
\hat{\phi}^{(s)}(g)=\hat{\phi}^{(s)}_{-}(g)+\hat{\phi}^{(s)}_{+}(g)
=\int\!\!\!\!\!\!\!\!\sum_{k\in I}
\hat{a}^{(s)}_k\psi^{(s)}_k(g) +{\hat{a}^{(s)\dag}_k}\bar{\psi}^{(s)}_k(g),
\ee
where $\hat{a}^{(s)}_k,{\hat{a}^{(s)\dag}_k}$ denote the Fourier coefficients of the
expansion of $\hat{\phi}^{(s)}$ in the basis $\{\psi^{(s)}_k\}_{k\in I}$.
The interested reader can consult Refs.
\cite{conforme,2ndgaq} for a more technical exposition of the subject
in the context of a Group Approach to Quantization formalism \cite{GAQ}.

The infinitesimal, second-quantized, counterpart of the finite action
(\ref{repre}) can be written as:
\bea
\left[\hat{X}_\alpha,\hat{\phi}^{(s)}(g)\right]&=& \hbar
X_\alpha^R\hat{\phi}^{(s)}(g)\,,\nn\\
\left[\hat{X}_\alpha,\hat{a}^{(s)}_l\right]&=&\hbar\int\!\!\!\!\!\!\!\!\sum_k\left.
\frac{\partial\rho^{(s)}_{lk}(g)}{\partial g^\alpha}\right|_{g=e} \hat{a}^{(s)}_k\,,
\label{semid}
\eea
where $\rho^{(s)}_{lk}(g)\equiv \langle
\psi^{(s)}_l|\rho(g)|\psi^{(s)}_k\rangle$ are the matrix elements of $\rho$ in the basis
of ${\cal H}_s(G)$. The operators $\hat{X}_\alpha$ are the
second-quantized version of the infinitesimal (right-invariant)
generators
\be
X^R_\alpha\equiv R_\alpha^\beta(g)
\frac{\partial}{\partial g^\beta}\,,\;\;\; R_\alpha^\beta(g)=
\left.\frac{\partial (g'\bullet g)^\beta}{\partial
g'^\alpha}\right|_{g'=e}\label{derecho}
\ee
of the finite left action (\ref{repre}), which fulfil the
commutation relations:
\be
\left[\hat{X}_\alpha,\hat{X}_\beta\right]=\hbar
C^\gamma_{\alpha\beta}\hat{X}_\gamma\,,\;\;\;C^\gamma_{\alpha\beta}=
(R_\alpha^\sigma\partial_\sigma R^\kappa_\beta-R_\beta^\sigma\partial_\sigma
R^\kappa_\alpha)(R^{-1})^\gamma_\kappa\,,\label{gcal}
\ee
where $C^\gamma_{\alpha\beta}$ denote the structure constants of the
Lie-algebra of $G$. The commutator between fields in the Fourier space and
the configuration space is:
\bea
\left[\hat{a}^{(s)}_k,{\hat{a}^{(s)\dag}_l}\right]&=&\delta_{kl}\hat{1}\,,\nn\\
\left[\hat{\phi}^{(s)}_{-}(g),\hat{\phi}^{(s)}_{+}(g')\right]&=&
\Delta^{(s)}(g,g')\hat{1}\,,
\label{hei}
\eea
where
\be
\Delta^{(s)}(g,g')\equiv \langle g|g'\rangle=\int\!\!\!\!\!\!\!\!\sum_k
\psi^{(s)}_k(g)\bar{\psi}^{(s)}_k(g')\,\label{propagador}
\ee
denotes the propagator in the configuration space. In the last equality, a
closure relation $1=\int\!\!\!\!\!\!\!\!\sum_k\ket{\psi^{(s)}_k}\bra{\psi^{(s)}_k}$ has
been introduced. The Hilbert space of the second-quantized theory is the
representation space of the (infinite-dimensional) algebra of quantum
fields $\hat{\phi}$ and space-time symmetries $G$ with commutation
relations  (\ref{semid},\ref{gcal},\ref{hei}). It can be constructed
as the orbit of the creation operators acting on the vacuum:
\be
|N(k_1),...,N(k_j)\rangle\equiv
\frac{({\hat{a}^{(s)\dag}_{k_1}})^{N(k_1)}...({\hat{a}^{(s)\dag}_{k_j}})^{N(k_j)}}
{(N(k_1)!...N(k_j)!)^{1/2}}|0\rangle  \,,\label{estados}
\ee
where $N(k_j)$ means ``number of particles with quantum numbers  $k_j$''.
The vacuum $|0\rangle$ is characterized by the conditions:
\be
\hat{a}^{(s)}_k|0\rangle=0=\hat{X}_\alpha|0\rangle\,,\;\;\forall k\in I,\,
\alpha=1,\dots,{\rm dim}(G),
\ee
that is, it is annihilated by the destruction field operators $\hat{a}^{(s)}_k$
and it is invariant under the action of the basic symmetry group $G$
---i.e. it looks the same to any freely falling observer anywhere in the
homogeneous (curved) space $M\simeq G/P$. It can be seen that the
operators $\hat{X}_\alpha$ (e.g. the Hamiltonian) are written in terms
of the (basic) field operators as:
\be
\hat{X}_\alpha=\hbar\int_G{{d}^Lg\,\hat{\phi}^{(s)}_{+}(g)X^R_\alpha
\hat{\phi}^{(s)}_{-}(g)}\,.\label{nobasicos}
\ee

We shall see how the Lie algebra of the second-quantization group
$G^{(2)}=G\times_s(H-W)$ [the semidirect product of the basic
symmetry group $G$ times the infinite-dimensional Heisenberg-Weyl
group of fields $\phi$, that is, the solution manifold to the
field equations (\ref{fieldeq})] defined by the commutation
relations (\ref{semid},\ref{gcal},\ref{hei}) is just a piece of a
richer structure that incorporates the whole algebra of
diffeomorphisms of $M$, diff$(M)$, ---as an enlargement of $G$---
together with all the representations ${\cal H}_s(G)$
(``higher-$G$-spin fields'') of $G$, in a consistent manner. We
shall denote such algebraic structures by $\cL_{\infty}({\cal
G})$, which are a particular (classical) limit of more general
tensor operator algebras $\cL_{\vec{\rho}}({\cal G})$ (see e.g
\cite{Pope,Fradkin2}) already mentioned in Sec. \ref{wbrief}.
Higher-spin algebras \cite{Fradkin}, which are said to be a guide
principle to the (still unknown) ``{\cal M}-theory'', are also of
this kind. Not much is known in the literature about the explicit
form of these algebras (e.g.: labelling of basic generators,
structure constants, etc), except for the simple cases of
$G=SU(2)$ and $G=SL(2,\R)\simeq SU(1,1)/\Z_2$, for which
$w_\infty\simeq \cL_{\infty}(sl(2,\R))$.

Before entering more explicitly
in this ``promotion'' of quantum-field-theory algebras
to higher structures containing ``gravity modes'', let us work out
the simple example of quantum field theory on AdS$_2=SO(1,2)/SO(1,1)$ as
a second quantization on $G=SO(1,2)\simeq SL(2,\R)$. The 4+1-dimensional
case AdS$_5=SO(4,2)/SO(4,1)$ is left as an Appendix \ref{intgeom}, where
explicit expressions for holomorphic unitary irreducible
representations of $SU(2,2)$ and propagators are given. Vacuum radiation
phenomena are also discussed as a physical application of
QFT in AdS$_2$.

\subsection{QFT in $AdS_2=SO(1,2)/SO(1,1)$\label{2ndgaqsec2}}

In restricting to  $1+1D$, we find an apparent ambiguity in the
distinction between  AdS$_2$ and  dS$_2$ in (\ref{AdS}), as
$SO(1,2)\simeq SO(2,1)$. We shall see that AdS$_2$ and dS$_2$
correspond to different orbits of the same group $SU(1,1)\sim
SO(1,2)$ (except for discrete transformations) or, in other words,
different choices of time generator ---see later on Eq. (\ref{twotimes}).

A system of coordinates for
\be
SU(1,1)=\left\{ g= \left( \begin{array}{cc} u_1 & \bar{u}_2 \\ u_2 &
\bar{u}_1\end{array}
\right) ,u_i,\bar{u}_i \in \mathbb{C}/ \det(g)=|u_1|^2-|u_2|^2=1 \right\}\,,
\ee
can be the following:\footnote{see also Appendix \ref{intgeom} for the
4+1-dimensional $SU(2,2)$ case}
\bea
g^{(1)}=\heta\equiv\f{u_1}{|u_1|}, \;\;g^{(2)} =z\equiv\f{u_2}{u_1},\;\;
g^{(3)}=\bar{z}\equiv\f{\bar{u}_2}{\bar{u}_1}\,,\nn\\
\heta\in U(1),\;\; z, \bar{z} \in D_1\;,\label{ycero}
\eea
where $D_1$ denotes the open unit disk in the complex plane $\C$.

The group law $g''=g'\bullet g$ in these coordinates adopts the form:
\begin{eqnarray}
\heta'' &=&\frac{u_1''}{|u_1''|}=\heta\frac{\sqrt{\heta'^2+
\al\alc'}}{\sqrt{1+\heta'^2\al'\alc}}\,,  \nn \\
\al''&=&\frac{u_2''}{u_1''}=\frac{\al'\heta'^2+\al}{\heta'^2+\al\alc'}\,,
\label{law1} \\
{\alc}''&=&\frac{{\alc_2''}}{{\alc_1''}}=
\frac{\alc\heta'^{2}+{\alc}'}{1+\heta'^{2}\al'\alc}\;.  \nn
\end{eqnarray}
The  matrices $L_\alpha^\beta$ and $R_\alpha^\beta$ in
(\ref{izquierdo},\ref{derecho}) corresponding to the left- and
right-invariant vector fields are (row $\alpha$, column $\beta$):
\be
(L_\alpha^\beta)=
\left(\ba{ccc} \heta & 0 &0 \\ \heta^{-1}\alc/2 &\heta^{-2}(1-\al\alc)&0\\
 -\heta^3\al/2 &0& \heta^2({1-\al\alc})\ea\right)\,,\;\;
(R_\alpha^\beta)=
\left(\ba{ccc} \heta & -2\al &2\alc \\ -\heta\alc/2 &1& -\alc^2\\
 \heta\al/2 &-\al^2&1 \ea\right)\,.\label{lrivf}
\ee
From here, the structure constants (\ref{gcal}) and commutation
relations of the $su(1,1)$
Lie algebra are:
\be
\lt[ \hat{X}_\heta,\hat{X}_\al \rt] =2\hbar\hat{X}_\al,\,\,
\lt[ \hat{X}_\heta,\hat{X}_{\alc} \rt] =-2\hbar\hat{X}_{\alc},\,\,
\lt[ \hat{X}_\al,\hat{X}_{\alc} \rt] =\hbar\hat{X}_\heta,\label{su11com}
\ee
which coincide with the well known $[L_m,L_n]=(m-n)L_{m+n}$ after the identification
$L_1\equiv\hat{X}_{\alc}, L_{-1}\equiv -\hat{X}_\al, L_0\equiv\um\hat{X}_{\heta}$.
The Casimir operator in this basis of generators is given by:
\be
\hat{C}=\hat{X}_\heta^2+2\hat{X}_\al^2\hat{X}_{\alc}^2+
2\hat{X}_{\alc}^2\hat{X}_\al^2\,.\label{casimir}
\ee

In order to construct the Hilbert space of wave functions on
AdS$_2$ and dS$_2$, let us start from the regular representation, that is,
complex functions  $\psi:SU(1,1)\to \mathbb C$ defined on the whole group
$SU(1,1)$. The  AdS$_2$ and dS$_2$ spaces correspond to
two different integral manifolds characterized by the vector fields
(perpendicular to the manifold)
\footnote{We shall consider holomorphic representations for the AdS case}
\be
{\rm AdS}_2:\; X_{p_1}^L\equiv X_{\al}^L \,,\;\;
{\rm dS}_2:\; X_{p_2}^L\equiv X_\al^L+X_{\alc}^L +iX_\heta^L\,,
\ee
and related to two possible choices of time:
\be
X_{t_1}^L\equiv \frac{\omega}{2} X_\heta^L\,,\;\;X_{t_2}^L\equiv \frac{i\omega}{2}
(X_\al^L-X_{\alc}^L)\,,\label{twotimes}
\ee
compact $t_1\sim \frac{2i}{\omega}\ln\heta$ for AdS$_2$ and non
compact $t_2\sim \frac{2i}{\omega}(\al-\alc)$ for dS$_2$ ($\omega$
is a parameter with frequency dimensions). The vector fields $X_{p_j}^L$ constitute
{\it (complex) polarizations} of coadjoint orbits \cite{coadjoint1}
in the language of Geometric Quantization
(see, for instance, \cite{Woodhouse})

Thus, wave functions do not depend on the coordinates $p_j$
``perpendicular'' to the manifold, which means
\be
X_{p_1}^L\psi_{{{\rm AdS}}}=0\,,\;\;
X_{p_2}^L\psi_{{{\rm dS}}}=0\,.\label{perpend}
\ee
The equations of motion
\be
X_{t_1}^L\psi_{{{\rm AdS}}}=\omega s_1\psi_{{{\rm AdS}}}\,,\;\;
X_{t_2}^L\psi_{{{\rm dS}}}=\omega s_2\psi_{{{\rm dS}}}
\ee
[the $SU(1,1)$-spin labels  $s_1, s_2\in \mathbb{Z}/2$ are related to the
{\it zero-point energy} ---or the {\it curvature} of
space \cite{2ndgaq}---  and characterize
the corresponding representation], together with (\ref{perpend}), yield the
solution:
\be
\psi_{{{\rm AdS}}}^{(s_1)}=\heta^{2s_1}(1-\al\alc)^{s_1}\,
\varphi_1(\alc)\,,\;\;
\psi_{{{\rm dS}}}^{(s_2)}=
\frac{(1-\al\alc)^{s_2}}{(\al\heta^2-i)^{s_2}(\alc\heta^{-2}+i)^{s_2}}\,
\varphi_2(\gamma)\,,
\ee
where $\gamma\equiv \heta^{-2}\frac{\al\heta^2-i}{\alc\heta^{-2}+i}$
corresponds to the spatial (compact) $S^1\subset$dS$_2$ coordinate, and
\be
\varphi_1\equiv \sum_{n=0}^\infty a^{(s_1)}_n \alc^n,\;\;\;
 \varphi_2 \equiv \sum_{m=-\infty}^\infty b^{(s_2)}_m \gamma^m
\ee
are arbitrary, analytic functions of $\alc$ and $\gamma$,
respectively. Note that the (holomorphic) AdS$_2$ representation
has a vacuum, whereas the spectrum in dS$_2$ has no lower-bound.

Let us restrict hereafter, for simplicity, to the AdS$_2$ case. Given
the natural integration measure
\bea
\d^Lg&\equiv& \frac{-i}{(2\pi)^2}
\theta^{L(\heta)}\wedge \theta^{L(\al)}\wedge \theta^{L(\alc)}=
\frac{-i}{(2\pi)^2} \det(L_\alpha^\beta)^{-1}d\heta\wedge d{\Re}(\al)\wedge
 d{\Im}(\al)\nn\\
&=&\frac{-i}{(2\pi)^2}\frac{1}{(1-\al\alc)^2} \heta^{-1}d\heta \wedge
d{\Re}(\al)\wedge d{\Im}(\al)\,,
\eea
the scalar product of basic functions
$\tilde{\psi}_n^{(s)}\equiv\heta^{2s}(1-\al\alc)^{s}\alc^n$,
\be
\langle \tilde{\psi}_m^{(s)}|\tilde{\psi}_n^{(s)}\rangle=
\frac{\Gamma(m+1)\Gamma(2s-1)}{\Gamma(2s+m)}\delta_{m,n}\equiv N^{(s)}_m
\delta_{m,n}\,,
\ee
is finite for ``conformal-spin'' $s>1/2$; here
$\Gamma$ denotes the standard  gamma function ($\Gamma(q)=(q-1)!$, when
$q\in \N$). Thus, $B({\cal H}_s)=\{\psi_n^{(s)}\equiv
\frac{1}{\sqrt{N^{(s)}_m}}\tilde{\psi}_n^{(s)}\}$ is an orthonormal basis
of the Hilbert space ${\cal H}_s(SU(1,1))$.
The matrix elements  $\rho^{(s)}_{mn}(g)\equiv\langle
\psi^{(s)}_m|\rho(g)| \psi^{(s)}_n\rangle$ of the
representation (\ref{repre}) in this basis have the following form:
\bea
\rho^{(s)}_{mn}(g)
&=& \sqrt{\frac{N^{(s)}_m}{N^{(s)}_n}}
\sum^m_{q={ \max}(0,m-n)}
\left(\begin{array}{c}n\\ m-q\end{array}\right)
\left(\begin{array}{c}2s+n+q-1\\ q\end{array}\right)\nn\\
& &\times
(-1)^{n-m+q}\heta^{-2s-2n}z^q \bar{z}^{n-m+q}(1-z\bar{z})^s,\label{matrixel}
\eea
and are the main ingredient to construct the quantum field theory
in AdS$_2$ (second quantization on $SU(1,1)$). Given the
expansion in modes of a real field with conformal-spin $s$,
\be
\hat{\phi}^{(s)}=\sum_{n=0}^\infty \hat{a}^{(s)}_n\psi^{(s)}_n+
\hat{a}_n^{(s)\dag}\bar{\psi}^{(s)}_n\,,\label{campoads}
\ee
the general commutation relations (\ref{semid}) adopt the
particular form
\bea
\lt[ \hat{X}_\heta, \hat{a}^{(s)}_n\rt] &=& -2\hbar(s+n)\hat{a}^{(s)}_n\,,\nn\\
\lt[ \hat{X}_\al, \hat{a}^{(s)}_n\rt] &=&\hbar\sqrt{n(2s+n-1)}\,
\hat{a}^{(s)}_{n-1}\,,\label{semidsu11}\\
\lt[ \hat{X}_{\alc,} \hat{a}^{(s)}_n\rt] &=&-\hbar\sqrt{(n+1)(2s+n)}\,
\hat{a}^{(s)}_{n+1}\,,\nn
\eea
together with the corresponding conjugated expressions
($\hat{X}_\heta^\dag=\hat{X}_\heta,\, \hat{X}_\al^\dag=- \hat{X}_{\alc}$).

For completeness, we shall give the explicit expression of the
propagator (\ref{propagador}) in the present holomorphic picture of
AdS$_2$, which is easily calculated as:
\be
\Delta^{(s)}(g,g')=\sum^{\infty}_{n=0}
\bar{\psi}^{(s)}_n(g')\psi^{(s)}_n(g)
=(2s-1)(\heta\heta'^{-1})^{2s}\frac{(1-\al'{\alc}')^s(1-\al\alc)^s}{(1-
\al'\alc)^{2s}}\,.\label{propaholo}
\ee

Before passing on to the embedding of the second-quantization group
$G^{(2)}$ [with general Lie-algebra commutation relations
(\ref{semid},\ref{gcal},\ref{hei}) and, more particularly,
(\ref{semidsu11},\ref{su11com})] in a larger structure containing
the diffeomorphisms of the manifold and higher-spin fields, let us
consider briefly an interesting thermodynamic result in AdS.

\subsection{Vacuum radiation\label{2ndgaqsec3}}

It is straightforward to see from (\ref{nobasicos})
that the {\it total energy} operator
$\hat{E}_{{\rm tot}}\equiv\frac{\hbar\omega}{2}\hat{X}_\heta$ is
written in terms of the creation and annihilation operators as:
\be
\hat{E}_{{\rm tot}}=\hat{E}_0+\hat{E}=
 E_0\hat{N}+ \sum_{n=1}^\infty E_n
\hat{a}^{(s)\dag}_{n}\hat{a}^{(s)}_{n}\label{openergiatotal}
\ee
where $E_0\equiv\hbar\omega s$ is the zero-point energy,
$E_n\equiv \hbar\omega n$ and  $\hat{N}\equiv\sum_{n=0}^\infty
\hat{a}^{(s)\dag}_{n}\hat{a}^{(s)}_{n}$ represents the operator {\it number
of particles}.

In addition to the ordinary vacuum $|0\rangle$, there are other
states with null renormalized energy
$\hat{E}=\hat{E}_{{\rm tot}}-\hat{E}_0$. For example, coherent
states of {\it zero modes}:
\be
 |0\rangle_\vartheta \equiv e^{-\frac{1}{2}|\vartheta|^2}
e^{\vartheta\hat{a}^{(s)\dag}_{0}}|0\rangle   \,, \label{thetavacuum}
\ee
which verify that: $\hat{a}^{(s)}_{n\geq 1}|0\rangle_\vartheta=0$ and
$\hat{a}^{(s)}_{0}|0\rangle_\vartheta=
\vartheta |0\rangle_\vartheta$, that is, they are eigenstates of the
annihilation operator $\hat{a}^{(s)}_{0}$. Although they behave as
vacua from an energetic point of view, these states exhibit a thermal
response under space-time transformations (let us say, accelerations).
Indeed, according to the general expression (\ref{matrixel}), the
finite action
\be
\hat{a}^{(s)\dag}_{0}\rightarrow \hat{b}^{(s)\dag}_{0}=
\sum^\infty_{n=0}\bar{\rho}^{(s)}_{0n}(\heta=1,\bar{z}=0,z)
\hat{a}^{(s)\dag}_{n}=\sum^\infty_{n=0}
(-1)^n\sqrt{\frac{N^{(s)}_0}{N^{(s)}_n}} z^n\hat{a}^{(s)\dag}_{n}\,,
\label{aceleracion}
\ee
generated by the `creation' operator $\hat{X}_z=-L_{-1}$, leads to the following transformation
of the vacuum (\ref{thetavacuum}) (for simplicity, we shall restrict
ourselves to the case $\vartheta=1$)
\be
|0\rangle_1 \rightarrow |\Psi(z)\rangle_1\equiv
 e^{-\frac{1}{2}}e^{\hat{b}^{(s)\dag}_{0}}|0\rangle=  \sum^\infty_{q=0}
z^q\sum_{\begin{array}{c}
m_1,...,m_q :\\ \sum^q_{n=1}n m_n=q\end{array}}
\prod^q_{n=0}\frac{r_n^{m_n}}{m_n!}
\prod^q_{n=0}(\hat{a}^{(s)\dag}_{n})^{m_n}|0\rangle \,,\label{bath}
\ee
where $r_n \equiv (-1)^n\sqrt{\frac{N^{(s)}_0}{N^{(s)}_n}}$ and
$m_0\equiv0$. This state has been identified in \cite{conforme} as
a (Weyl-invariant) vacuum seen
from an accelerated frame ---see Eq. (\ref{temper})--- 
in the context of a 1+1D conformally invariant QFT. This
comparison is justified because, in 1+1D, the
(finite) conformal group $SO(2,2)\simeq SU(1,1)\times SU(1,1)$ (except
for discrete symmetries) splits into left- and right-moving modes, so
that we are just dealing with ``one direction'' when working
in AdS$_2$. The transition to an accelerated frame is realized by
special conformal transformations: $L_{-1}$ and $\bar{L}_{-1}$
(see \cite{conforme} for more details).

The relative probability of observing a state with total energy
$E_q=\hbar\omega q$ in the accelerated vacuum $|\Psi(z)\rangle_1$ is
\bea
P_q&=&\Lambda(E_q)(|z|^2)^q \,,\nn\\
\Lambda(E_q)&\equiv &\sum_{\begin{array}{c}
m_1,...,m_q :\\ \sum^q_{n=1}n m_n=q\end{array}}
\prod^q_{n=0}\frac{r_n^{2m_n}}{m_n!}\,.
\eea
It is interesting to see how we can associate a {\it thermal bath}
with this distribution function by noticing that $\Lambda(E_q)$ behaves
as a relative weight proportional to the number of states with energy
$E_q=\hbar\omega q$; the factor $(|z|^2)^q$ fits this weight properly to a
temperature as
\be
(|z|^2)^q=e^{q\log|z|^2}=e^{-\frac{E_q}{k_BT}}\,,\,\,\,\;
\hbox{where}\,\,\,\,\; T\equiv-\frac{\hbar\omega}{k_B\log|z|^2}=\frac{\hbar  a}
{2\pi ck_B} \label{temper}
\ee
is the temperature associated with a given acceleration
$a\equiv \frac{2\pi\omega c}{\log|z|^2}$, and $k_B,c$ denote the
Boltzmann constant and speed of light, respectively. This simple, but
profound, result was first considered by Unruh
\cite{Unruh}.

After some intermediate calculations, the expected value
of the total energy $\hat{E}$ in the accelerated vacuum $|\Psi(z)\rangle_1$
proves to be:
\be
\frac{{}_1\langle\Psi(z)|\hat{E}|
\Psi(z)\rangle_1}{{}_1\langle\Psi(z)|
\Psi(z)\rangle_1}=s\hbar\omega\frac{|z|^2}{(1-|z|^2)^{2s+1}}\,,\label{emedia}
\ee
which coincides with the {\it mean energy per mode} of the
Bose-Einstein statistic, for $s=0$, when we substitute
(\ref{temper}) inside (\ref{emedia}). In $D$ spatial dimensions,
the number of states with frequency $\omega$ is proportional
to $\omega^{D-1}$. Thus, the spectral distribution of the radiation
of the accelerated vacuum $|\Psi(z)\rangle_1$ for $D=3$ is given by the
formula
\be
u_s(x)=\epsilon_0\frac{x^3 e^{-x}}{(1-e^{-x})^{2s+1}}\,, \ee where
$x\equiv\frac{\hbar\omega}{k_BT_0}$ and $\epsilon_0$ is a
constant, for a fixed temperature $T_0$,  with dimensions of
energy per unit of volume. Figure \ref{Planck} represents the
spectral distribution $u_s(x)$ for different values of $s$
(related to physical magnitude like the zero-point energy and the
curvature of the space). For $s\to 0$, we recover the Planckian
(black body) spectrum.\footnote{Note that this limit can only be
reached in going to the universal covering of $SU(1,1)$, which
means to make time non-compact.} For $s>0$, we have a deformation
of the Planckian spectrum; the value $s=(D-1)/2$ corresponds to a
{\it critical} situation: over this value, the theory exhibits an
``infrared catastrophe''. The physical meaning of this divergence
is unclear to the author. A possible explanation could be that
higher-spin fields exhibit anomalies and require  more careful
treatment. The reason for this belief lies in the fact that
quantization deforms higher-spin algebras like $w$ to $\W$, by
introducing renormalizations ---Moyal terms--- (see next Section).

\begin{figure}[htb]
\begin{center}
\includegraphics[height=6cm,width=10cm]{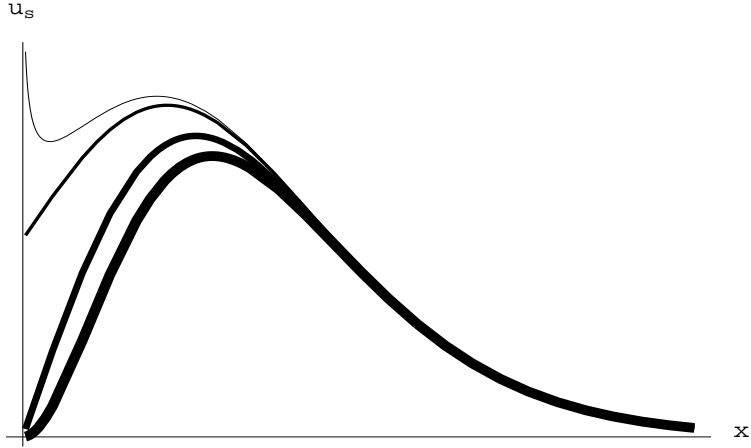}
\caption{{ {\sf Higher-spin $s=\frac{1}{2},1$({\rm critical}),$>1$(infrared
divergence) deformations of the Planckian spectrum $s=0$
(thickest curve).}}} \label{Planck}
\end{center}
\end{figure}

Unruh effect could be seen as a ``first order effect'' of gravity
on quantum field theory, in the sense that uniform accelerations
(special conformal transformations) are enough to account for this
radiation phenomenon. Higher order (linear, quadratic, etc) accelerations
can be consistently included in 1+1D by enlarging the $SO(2,2)
\sim SU(1,1)\times SU(1,1)$ symmetry to two copies of the
Virasoro algebra
\bea
\leftarrow\dots L_{-3},L_{-2},&\overbrace{L_{-1},L_0,L_1}^{su(1,1)}&,
L_2,L_3\dots\rightarrow \nn\\
\leftarrow\dots \bar{L}_{-3},\bar{L}_{-2},&\underbrace{\bar{L}_{-1},
\bar{L}_0,\bar{L}_1}_{su(1,1)}&,
\bar{L}_2,\bar{L}_3\dots\rightarrow, \label{alargamiento}
\eea
for the generators $L_n, \bar{L}_n$ of holomorphic and antiholomorphic
diffeomorphisms (left- and right- moving modes), respectively, with commutation relations:
\bea
\left[L_m,L_n\right]&=&(m-n)L_{m+n}\nn+\frac{c}{12}m(m^2-1)\delta_{m+n,0}\nn\\
\left[\bar{L}_m,\bar{L}_n\right]&=&(m-n)\bar{L}_{m+n}
+\frac{c}{12}m(m^2-1)\delta_{m+n,0}\label{2vira}\\
\left[L_m,\bar{L}_n\right]&=&0,\nn
\eea
where $c$ denotes the typical central charge.

We have given a statistical mechanical description of the
Unruh effect (vacuum radiation in uniformly accelerated frames)
just by counting microscopic states of a conformally invariant
quantum field theory ---more precisely, the restriction to its
left-(or right-)moving sector.  There is strong evidence
(see \cite{Carlip9812013} and references therein)
that conformal field theories also provide a universal description of
low-energy black hole thermodynamics. Actually, Unruh's temperature
(\ref{temper}) coincides with Hawking's temperature
\begin{equation}
T= \frac{\hbar c^3}{8\pi M k_B G}=\frac{2\pi GM\hbar}{\Sigma c k_B}
\label{tempHawking}
\end{equation}
($\Sigma=4\pi r_g^2=
8\pi G^2M^2/c^4$ stands for the surface of the event horizon)
when the acceleration is that of a free falling observer on the surface
$\Sigma$, i.e.  $a=c^4/(4GM)=GM/r_g^2$. In fact,
the Virasoro algebra (\ref{2vira}) proves to be a physically important
subalgebra of the gauge algebra of surface deformations that leave
the horizon fixed for an arbitrary black hole. Thus, the fields on the surface
must transform according to irreducible representations of the Virasoro
algebra (\ref{2vira}), which is the general symmetry principle that
governs the density of microscopic states. Bekenstein-Hawking expression
for the entropy can be then calculated from the Cardy formula
\cite{Cardy} (see also \cite{Carlip} for logarithmic corrections).

The infinite-dimensional character of conformal symmetry seems to be
exclusive patrimony of 1+1 dimensions. Indeed, conformal symmetry
$SO(D,2)$ in $D$ space-time dimensions is said to be finite-dimensional
except for $D=2$. However, one can consider ``analytic continuations''
(in a broad sense specified in the next Section)
of the finite-dimensional $SO(4,2)$ symmetry, containing
diffeomorphisms and higher-spin fields, which constitutes an extrapolation
of the Virasoro and $\W$ symmetries to realistic (3+1) dimensions. Such
infinite enlargement of $SO(4,2)\sim SU(2,2)$ has been given by the author
\cite{infdimal,v3p1} in the (more general)
context of a new class of infinite-dimensional tensor
operator algebras of pseudo-unitary groups $U(N_+,N_-)$. Let us
see how these abstract algebras admit a physical interpretation as
Lie algebras of higher-spin fields and diffeomorphisms on
homogeneous spaces like AdS$_2$ and AdS$_5$.

\section{Higher-dimensional $\W$ algebras and
quantum AdS$_5$ space.\label{v3p1}}

The second quantization algebra (\ref{semid},\ref{gcal},\ref{hei})
contains the strictly necessary space-time symmetries (\ref{gcal}) from
which to extract the space-time manifold. Nevertheless, in (1+1) dimensions,
the $su(1,1)$ Lie algebra (\ref{su11com}) can be enlarged
to the Virasoro algebra (\ref{2vira}) through an ``analytic continuation''
(\ref{alargamiento}), that is, an extension of $su(1,1)=\{L_n,|n|\leq 1\}$
beyond the wedge $|n|\leq 1$ when we identify
$\hat{L}_1\equiv \hat{X}_{\alc},\,\hat{L}_{-1}\equiv -\hat{X}_{\al},
\hat{L}_0\equiv\frac{1}{2}\hat{X}_{\heta}$. This incorporation of new
local transformations (diffeomorphisms) constitutes the main (gauge) guide
principle to formulate two-dimensional gravity models, like the
well known Polyakov's induced gravity \cite{Poly}. There can be no doubt
that a similar enlargement procedure promoting the finite-dimensional
$SO(4,2)$ conformal symmetry to the infinite realm is a potentially
valuable symmetry resource to formulate gravity models in realistic
dimensions. Let us show how this infinite enlargement of
$SO(4,2)\sim U(2,2)/U(1)$ can be given as a particular case of more
general infinite extensions of pseudo-unitary symmetries $U(N_+,N_-)$,
and how diffeomorphisms and higher-spin currents are embedded in this
algebraic structure.

Let us denote by ${X}_{\alpha\beta},\, \alpha,\beta=1,\dots,N\equiv N_++N_-$,
the generators (step operators) of the $u(N_+,N_-)$ Lie algebra
(see Appendix \ref{intgeom} for another basis of
generators) , which admit a matrix realization of the form:
\be
X_{\alpha\beta}\equiv -{X_\alpha}^\gamma\eta_{\gamma\beta},\;\;{\rm with}\;\;
{({X_{\alpha}}^{\gamma})_{\sigma}}^{\beta}\equiv
\delta_{\alpha\sigma}\delta^{\gamma\beta},\label{step}
\ee
where  $\eta={\rm diag}(1,\stackrel{N_+}{\dots},1,-1,
\stackrel{N_-}{\dots},-1)$ is used to raise and lower indices.
The commutation relations of these step operators are:
\be
\left[{X}_{\alpha_1\beta_1},{X}_{\alpha_2\beta_2}\right]=
(\eta_{\alpha_1\beta_2}{X}_{\alpha_2\beta_1}-
\eta_{\alpha_2\beta_1}{X}_{\alpha_1\beta_2})\,.\label{pun}
\ee
The relation between the step operators ${X}_{\alpha\beta}$ and $L_n$ for
$u(1,1)$ is simply: $L_1=X_{12}, L_{-1}=X_{21},
L_0=\frac{1}{2}(X_{22}+X_{11}), {\cal N}=\frac{1}{2}(X_{22}-X_{11})$, where
${\cal N}$ is the generator associated with the trace
$\eta_{\alpha\beta}X^{\alpha\beta}$ ---the operator {\it number of particles}
(\ref{openergiatotal}) or {\it electric charge} in second quantization. The key point is to
note that the commutation relations (\ref{pun}) can also be written
as:
\be
\left[L^I_m, L^J_n\right]=\eta^{\alpha\beta}(J_\alpha m_\beta-
I_\alpha n_\beta)L^{I+J-\delta_\alpha}_{m+n}\,,\label{inf}
\ee
where the lower index  $m$ of $L$ now symbolizes an integral upper-triangular
$N\times N$ matrix and the upper $U(N_+,N_-)$-spin index $I$
represents a half-integral lattice vector; more schematic:
\be
m=\left(\ba{ccccc} 0 & m_{12}& m_{13} & \dots & m_{1N}\\
0 & 0 & m_{23} & \dots & m_{2N} \\ 0 & 0 & 0 & \dots & m_{3N} \\
 \vdots & \vdots & \vdots & \ddots  & \vdots \\
\vdots & \vdots & \vdots & \ddots  & 0 \ea \right)_{N\times N},
m_{\alpha\beta}\in \Z\,;\;\;\;I=(I_1,\dots, I_N)\,,\;\;I_\alpha\in \Z/2\,.
\label{uppertmatrix}
\ee
We are also denoting $m_\alpha\equiv
\sum_{\beta>\alpha} m_{\alpha\beta}-\sum_{\beta<\alpha} m_{\beta\alpha}$ and
$\delta_\alpha\equiv (\delta_{\alpha,1},\dots,\delta_{\alpha,N})$.
Note that the generators ${L}^I_m$ are labeled by $N+N(N-1)/2=N(N+1)/2$
indices, in the same way as wave functions $\psi^I_m$ in the
carrier space of irreps of $U(N)$ (see Appendix \ref{intgeom}).

There are many possible ways of embedding the $u(N_+,N_-)$ generators
(\ref{pun}) inside (\ref{inf}), as there are many possible choices
of $su(1,1)$ inside (\ref{2vira}). However, we can establish a
``canonical'' choice, such as:
\be
X_{\alpha\beta}\equiv
L^{\delta_\alpha}_{x_{\alpha\beta}}\,, \;\;\;
x_{\alpha\beta}\equiv {\rm sign}(\beta-\alpha)
\sum_{\sigma={{\min}}(\alpha,\beta)}^{{{\max}}(\alpha,\beta)-1} x_{\sigma,\sigma+1}\,,\label{embedding}
\ee
where $x_{\sigma,\sigma+1}$  denotes an upper-triangular matrix
with the $(\sigma,\sigma+1)$ entry equal to one and zero elsewhere, 
i.e. $(x_{\sigma,\sigma+1})_{\mu\nu}=
\delta_{\sigma,\mu}\delta_{\sigma+1,\nu}$.
For example, the $u(1,1)$ generators correspond to:
\be
X_{12}=L^{(1,0)}_{\left(\begin{array}{cc} 0 &1\\ 0&0\ea\right)},\;\;
X_{21}=L^{(0,1)}_{\left(\begin{array}{cc} 0 &-1\\ 0&0\ea\right)},\;\;
X_{11}=L^{(1,0)}_{\left(\begin{array}{cc} 0 &0\\ 0&0\ea\right)},\;\;
X_{22}=L^{(0,1)}_{\left(\begin{array}{cc} 0 &0\\ 0&0\ea\right)}.
\ee

If we allow the lower-index $m=x_{\alpha\beta}$ of $L^{\delta_\alpha}_m$ in
Eq. (\ref{embedding}) to run over arbitrary integral upper-triangular
matrices $m$, then we arrive at the infinite-dimensional algebra:
\be
\lt[ L^{\delta_\alpha}_{m},
L^{\delta_\beta}_{n}\rt] = m^\beta L^{\delta_\alpha}_{m+n}-
n^\alpha L^{\delta_\beta}_{m+n}\,, \label{difeounm}
\ee
which we shall denote by $\cL_{\infty}^{(1)}(u(N_+,N_-))$. As a particular
example, for the simplest case of $u(1,1)$, this
``analytic continuation'' leads to two Virasoro
sectors: $L_{m_{12}}\equiv L^{(1,0)}_m,\,
\bar{L}_{m_{12}}\equiv L^{(0,1)}_m$.
Its (3+1)-dimensional counterpart
$\cL_{\infty}^{(1)}(u(2,2))$ contains four noncommuting
Virasoro-like sectors
$\cL_{\infty}^{(1_\alpha)}(u(2,2))=\{L^{\delta_\alpha}_{m}\}
,\,\alpha=1,\dots,4$, which, in their turn, hold three genuine Virasoro
sectors for  $m=k u_{\alpha\beta},\,
k\in \Z,\, \alpha,\beta=1,\dots,4$, where $(u_{\alpha\beta})_{\mu\nu}=
\delta_{\alpha,\mu}\delta_{\beta,\nu}$ is an upper-triangular matrix.
In general,
$\cL_{\infty}^{(1)}(u(N_+,N_-))$ contains $N(N-1)$ distinct and noncommuting
Virasoro sectors, and holds $u(N_+,N_-)$ as the {\it maximal
finite-dimensional subalgebra}.

The algebra $\cL_{\infty}^{(1)}(u(N_+,N_-))$ can be seen as the {\it minimal}
infinite continuation of $u(N_+,N_-)$ representing the diffeomorphism
algebra  diff$(N)$ of the corresponding $N$-dimensional manifold (locally the
Minkowski space-time for $u(2,2)$). Indeed, the algebra (\ref{difeounm})
formally coincides with the algebra of vector fields
$L^\mu_{f(y)}=f(y)\frac{\partial}{\partial y_\mu}$, where
$y=(y_1,\dots,y_N)$ denotes a local system of coordinates and $f(y)$
can be expanded in a plane wave basis, such that
$L^\mu_{\vec{m}}=e^{im^\alpha y_\alpha}
\frac{\partial}{\partial y_\mu}$
constitutes a basis of vector fields for  the so called
generalized Witt algebra
\cite{Ree}, of which there are studies about its representations (see e.g.
\cite{FigueiridoRamos,Rao,Fabbri,Larsson99}). Note that, for us, the $N$-dimensional
lattice vector $\vec{m}=(m_1,\dots,m_N)$ is constrained by
$\sum_{\alpha=1}^N m_\alpha=0$ (see the definition of $m_\alpha$ in
paragraph after Eq. \ref{inf}), which
introduces some novelties as regards the Witt algebra. In fact, the algebra
(\ref{difeounm}) can be split into one ``temporal'' piece, constituted by
an Abelian ideal generated by $\tilde{L}^N_m\equiv \eta_{\alpha\alpha}
L^{\delta_\alpha}_{m}$, and a ``residual'' symmetry generated by the
spatial diffeomorphisms
\be
\tilde{L}^j_m\equiv\eta_{jj}
L^{\delta_j}_{m}-\eta_{j+1,j+1} L^{\delta_{j+1}}_{m},\,j=1,\dots,N-1\,\,
({\rm no \ sum \ on \ } j)\,,
\ee
which act semi-directly on the temporal part. More precisely, the
commutation relations (\ref{difeounm}) in this new basis adopt the following
form:
\bea
\l[ \tilde{L}^j_m,\tilde{L}^k_n\r] &=& \tilde{m}^k \tilde{L}^j_{m+n} -
\tilde{n}^j \tilde{L}^k_{m+n}\,,\nn\\
\l[ \tilde{L}^j_m,\tilde{L}^N_n\r] &=&  -\tilde{n}^j \tilde{L}^N_{m+n}\,,
\label{inftempesp}\\
\l[ \tilde{L}^N_m,\tilde{L}^N_n\r] &=& 0\,,\nn
\eea
where $\tilde{m}_k\equiv m_k-m_{k+1}$. Only for $N=2$, the last
commutator admits a central extension of the form
$\sim n_{12}\delta_{m+n,0}$
compatible with the rest of commutation relations
(\ref{inftempesp}). This result amounts to the fact that the
(unconstrained) diffeomorphism algebra diff$(N)$ does not admit any
non-trivial central extension except when $N=1$ \cite{nocentral}.

Additionally, after the restriction
$I=\delta_\alpha$ in (\ref{embedding}) is also relaxed to arbitrary
half-integral lattice vectors $I$ (higher-$U(N_+,N_-)$-spins), the commutation
relations (\ref{inf}) define a {\it
higher-$u(N_+,N_-)$-spin algebra} $\cL_{\infty}(u(N_+,N_-))$ (in a sense
similar to  that of Ref. \cite{Fradkin}), which
contains $\cL_{\infty}^{(1)}(u(N_+,N_-))$ as a subalgebra as well as
all (interacting) {\it currents} $L^I_m$ with all $U(N_+,N_-)$-spins $I$.

For example, for the conformal-spin $s$ quantum fields $\hat{\phi}^{(s)}$
in (\ref{campoads}) on AdS$_2= SO(1,2)/ SO(1,1)$, one can
check that the identification:
\bea
L^{(-s,0)}_{|m|}=\sqrt{N^{(s)}_{|m|-s}}\hat{a}^{(s)\dag}_{|m|-s}\,,\;\;\;\;
L^{(-s,0)}_{-|m|}=\sqrt{N^{(s)}_{|m|-s}}\hat{a}^{(s)}_{|m|-s}\,,\nn\\
L^{(1,0)}_{\left(\ba{cc} 0 & 1 \\ 0 & 0\ea\right)}=\hat{X}_z\,,\;\;
L^{(1,0)}_{\left(\ba{cc} 0 & -1 \\ 0 & 0\ea\right)}=-\hat{X}_{\bar{z}}\,,\;\;
L^{(1,0)}_{\left(\ba{cc} 0 & 0 \\ 0 & 0\ea\right)}=-\frac{1}{2}\hat{X}_{\heta}
\eea
between the $\cL_{\infty}(u(1,1))$ generators
$L^{(I_1,I_2)}_m$, the space-time $SU(1,1)$ generators (\ref{su11com})
and the Fourier coefficients $\hat{a}^{(s)}_m,\hat{a}^{(s)\dag}_m$ of the
field $\hat{\phi}^{(s)}$
---where $|m|$ denotes the absolute value of all the entries of the upper-triangular matrix
$m=\left(\ba{cc} 0 & m \\ 0 & 0\ea\right)$---, give us an embedding
of the second quantization algebra (\ref{su11com},\ref{semidsu11}) in (\ref{inf}),
except for central terms. Thus, the algebra
$\cL_{\infty}(u(1,1))$ contains not just the basic space-time symmetry
$u(1,1)$ of AdS$_2$, but also its prolongation to the whole algebra
$\cL_{\infty}^{(1)}(u(1,1))\sim {\rm diff}(2)$ of diffeomorphisms
of AdS$_2$, together with all fields $\hat{\phi}^{(s)}$ with
arbitrary conformal spin. This embedding of $SU(1,1)$-spin $s$ quantum fields
$\hat{\phi}^{(s)}$ on AdS$_2$ in $\cL_{\infty}(u(1,1))$
is straightforwardly generalized to
$SU(2,2)$-spin $S$ quantum fields $\hat{\phi}^{(S)}$ on AdS$_5$ (see
Appendix \ref{intgeom} for a explicit expression of coordinates, wave
functions and propagators in AdS$_5$) by considering the
(4+1)-dimensional counterpart $\cL_{\infty}(u(2,2))$ of the
(1+1)-dimensional higher-spin algebra $\cL_{\infty}(u(1,1))$.
Note that the generators $L^{I}_m$ of $\cL_{\infty}(u(N_+,N_-))$
carry an extra spin label (the trace $S_N=I_1+\dots+ I_N$) with respect
to the Fourier coefficients $a^{(S)}_m$, $S=(S_1,\dots,S_{N-1})$, of the
$SU(N_+,N_-)$ quantum fields $\hat{\phi}^{(S)}$ [see paragraph before
Eq. (\ref{cartanact})]; the relevance of this
extra phase invariance and its possible connection with charge conjugation
remains to be studied.

As we have already said, the algebra (\ref{inf}) reproduces the
commutation relations of higher-$U(N_+,N_-)$-spin quantum
fields except for central terms (propagators).
In fact, besides such central extensions, the quantization process
entails unavoidable renormalizations (mainly due to ordering problems)
of the form:
\be
\left[\hat{L}^I_m, \hat{L}^J_n\right]=
\hbar\eta^{\alpha\beta}(J_\alpha m_\beta-
I_\alpha n_\beta)\hat{L}^{I+J-\delta_\alpha}_{m+n}+ O(\hbar^3) +
\hbar^{(\sum_{\alpha=1}^N{I_\alpha+J_\alpha})}c^{(I,J)}(m)\delta_{m+n,0}\hat{1}
\,,\label{infq}
\ee
where $\hat{1}\sim \hat{L}^0_0$ denotes a central generator and
$c^{(I,J)}(m)$ are central charges. The higher order terms $O(\hbar^3)$
can be captured in a classical construction by extending the
classical (Poisson-Lie) bracket (\ref{inf}) to the Moyal bracket (see
\cite{infdimal} for more information on Moyal deformation).

Central extensions provide the essential ingredient
required to construct invariant geometric
action functionals on coadjoint orbits of the corresponding groups.
When applied to the infinite continuation
(\ref{infq}) of $u(2,2)$, this would lead
to Wess-Zumino-Witten-like models for {\it induced conformal
gravities in $3+1$ dimensions}, as  happens for the Virasoro and
${\cal W}$ algebras in relation with (1+1)-dimensional gravity
(see e.g. Ref. \cite{Nissimov}). We just claim here the potential (gauge) guiding
principle of the higher-$u(2,2)$-spin algebra (\ref{infq}) towards
the formulation of consistent gravity models in realistic dimensions. The 
explicit construction of these models is left for future work.

Moreover, as we have already commented, the statistical mechanical
explanation of radiation phenomena like black hole evaporation
proves to rely heavily on the count of microscopic states of a
given gauge (diffeomorphism) invariant field theory. Previous to
this, an exhaustive study of central charges and representation
theory of the corresponding symmetry algebras ---like
(\ref{infq})--- would be necessary. The general study of
infinite-dimensional algebras and groups has not progressed very
far, except for some important achievements in one and two
dimensions. Thus, we do not expect the representation theory of
these algebras to be an easy task. Perhaps, some interesting results imported
from the representation theory of the algebra of diffeomorphisms of the $n$-torus 
\cite{FigueiridoRamos} and toroidal Lie algebras (see e.g.
\cite{Billing}) could be of help. ``Two-toroidal Lie algebras''
are higher-dimensional generalizations of the affine Kac-Moody
algebra in the sense that they replace the loop with a two-torus,
that is, the integer lower-index $m$ (loop winding number) of the
Kac-Moody algebra generator $T^a_m$ is replaced by an integral
four-vector $\vec{m}=(m_1,\dots,m_4)$. These algebras appear as
current symmetry algebras of the four-dimensional K\"ahler WZW
model, also denoted by WZW$_4$ (see e.g. \cite{Inami}). WZW$_4$
model also arises as the induced surface theory of a 4+1
dimensional AdS Chern-Simons gravity theory with gauge group
$SO(4,2)\times U(1)\sim U(2,2)$ ---except for discrete
symmetries--- (see \cite{Banados,Gegenberg}), in analogy with what
happens in 2+1 dimensions \cite{Achucarro,Witten}.

\subsection{Quantum AdS space\label{qadse}}

Before passing on to the Conclusions, let us comment on the potential
relevance of the $C^*$-algebras (\ref{infq}) on tractable non-commutative
versions of curved spaces like AdS$_5$. It is a remarkable fact that
the algebra (\ref{infq}) is actually a member
of a $N$-parameter family $\tilde{\cL}_{\vec{\rho}}(u(N_+,N_-)),\,
\vec{\rho}\equiv (\rho_1,\dots,\rho_N)$ of non-isomorphic algebras of
$U(N_+,N_-)$ tensor operators (see \cite{infdimal}),
the classical limit $\hbar\to 0, \rho_\alpha\to \infty$  corresponding to
the classical (Poisson-Lie) algebra $\cL_{\infty}(u(N_+,N_-))$ with
commutation relations (\ref{inf}). A very interesting feature of
$\tilde{\cL}_{\vec{\rho}}(u(N_+,N_-))$ is that it {\it collapses} to
${\rm Mat}_{d}({\C})$ (the full matrix algebra of $d\times d$
complex matrices) whenever the (complex) parameters
$\rho_\alpha$ coincide with the eigenvalues $q_\alpha$
of the Casimir operators $C_\alpha$ of $u(N_+,N_-)$ in a
$d$-dimensional irrep $D_{\vec{q}}$ of
$u(N_+,N_-)$. This fact can provide discrete,
`fuzzy' or `cellular' descriptions of the non-commutative counterpart
of AdS$_5$ when applying the ideas of {\it non-commutative geometry} (see e.g.
\cite{Connes}) to $\tilde{\cL}_{\vec{\rho}}(u(2,2))$.  The appealing feature
of a non-commutative space $M$ is that a $G$-invariant `lattice structure'
can be constructed in a natural way, a desirable property as regards
finite models of quantum gravity (see e.g. \cite{Madore} and Refs. therein).

It is also a very important feature of
$\cL_{\vec{\rho}}(u(N_+,N_-))$ that the quantization deformation scheme
(\ref{Moyaleq}) does not affect the maximal finite-dimensional subalgebra
$su(N_+,N_-)$ (`good observables' or preferred coordinates \cite{Bayen}) of
non-commuting `position operators'
\bea
&y_{\alpha\beta}=\frac{\kbar}{2}({X}_{\alpha\beta}
+{X}_{\alpha\beta}^\dag)\,,\;\;
y_{\beta\alpha}=\frac{i\kbar}{2}({X}_{\alpha\beta}
-{X}_{\beta\alpha}^\dag)\,,\;\;\;\;\alpha<\beta\,,&\nn\\
&y_\alpha={\kbar}(\eta_{\alpha\alpha}{X}_{\alpha\alpha}-
\eta_{\alpha+1,\alpha+1}{X}_{\alpha+1,\alpha+1})\,,&
\eea
on the algebraic manifold $F_{{}_{N_+N_-}}$, where $\kbar$
gives $y$ dimensions of length.
The `volume' $v_j$ of the
$N-1$ submanifolds $F_j$ of the {\it flag manifold} $F_{{}_{N_+N_-}}=
F_{N}\supset\dots\supset F_2$ (see e.g. \cite{Helgason,Fulton} for a definition
of flag manifolds)
is proportional to the eigenvalue
$\rho_j$ of the $su(N_+,N_-)$ Casimir operator $\hat{C}_{j}$ (\ref{casisun})
in those coordinates:
$v_j=\kbar^{j}\rho_j$. Large volumes (flat-like spaces) correspond to
a high density of quantum points (large $\rho$). In the classical limit
$\kbar\to 0$, $\rho\to \infty$, the coordinates $y$ commute.

The notion of ``quantum space" itself could be the collection
of all of them, enclosed in a more general algebraic structure. A suitable candidate for this purpose,
where different quantum spaces can coexist with different multiplicities,
is the {\it group algebra} $C^*(G)$ of complex functions on $G$. Indeed, the
group algebra decomposes as (see e.g. \cite{coadjoint1,Landsmanbook}):
\be
C^*(G)\simeq \bigoplus_{\vec{\rho}\in\hat{G}}\cL_{\vec{\rho}}({\cal G}),
\ee
where $\hat{G}$ denotes the set of unirreps of $G$. From this point of
view, classical, flat-like spaces (large $\rho$) would be ``more likely"
to be found inside the whole ensemble $C^*(G)$, as it should be!.

This breaking down of $C^*(G)$
into factor algebras $\cL_{\vec{\rho}}({\cal G})$
(matrix algebras for $\vec{\rho}\in \hat{G}$) is the
quantum analogue of the standard foliation ${\cal G}^*\simeq \bigcup_j F_j$
of the coalgebra ${\cal G}^*$ into coadjoint orbits $F_j$ (symplectic leaves)
\cite{Landsmanbook}.

There is a natural invariant associative and
noncommutative $*$-product between functions
$\psi^I_m,\psi^J_n\in C^*(G)$ (the {\it convolution product}):
\be
(\psi^I_m *\psi^J_n)(g')\equiv\int_G d^Lg\,\psi^I_m(g)
\psi^J_n(g^{-1}\bullet g'), \label{gaconvo}
\ee
and an {\it involution} $\psi^*(g)\equiv\bar{\psi}(g^{-1 })$, which define $C^*(G)$ as
a noncommutative $C^*$-algebra. We also define the ``negative modes" $\psi^I_{-m}$ of
basic anti-holomorphic functions (\ref{antiholf},\ref{vacsu22}) as the corresponding
holomorphic ones, i.e. $\psi^I_{-m} \equiv\bar{\psi}^I_m$.
The classical limit of the convolution commutator
\be
\{\psi,\psi'\}=
\lim_{\kbar\to 0}\frac{i}{\kbar^2}(\psi *\psi'-\psi' *\psi)
\ee
corresponds to the Poisson-Lie bracket between functions on $C^\infty({\cal G}^*)$.
The explicit form (in coordinates) of the Poisson-Lie bracket for $G=U(N_+,N_-)$ is:
\be
\left\{\psi,\psi'\right\}=
(\eta_{\alpha_1\beta_2}
{x}_{\alpha_2\beta_1}-\eta_{\alpha_2\beta_1}{x}_{\alpha_1\beta_2})
\frac{\partial \psi}{\partial x_{\alpha_1\beta_1}}
\frac{\partial \psi'}{\partial x_{\alpha_2\beta_2}},\label{poissonlie}
\ee
where $x_{\alpha\beta}, \alpha,\beta=1,\dots,N$  denote the coordinates in the coalgebra
${\cal G}^*$ (seen as a $N^2$-dimensional vector space) of $G=U(N_+,N_-)$.
The constraints $C_\alpha(x)=\kbar^\alpha\rho_\alpha\equiv v_\alpha$ induced by the
Casimir operators (\ref{casisun}) foliate $C^\infty({\cal G}^*)$ into
Poisson algebras $C^\infty(F_\alpha)$. The leaves $F$
(coadjoint orbits, flag manifolds) of ${\cal G}^*$ admit a symplectic
structure  $(F,\Omega)$, where $\Omega$ denotes
a closed two-form ---K\"ahler form (\ref{kahlerform})--- which can be obtained from
a K\"ahler potential (\ref{kahlerpot}) as described at the end of Appendix
\ref{intgeom}. After this foliation, the Poisson bracket (\ref{poissonlie}) becomes
\be
\left\{\psi_a,\psi_b\right\}=\sum_{\alpha_j>\beta_j}
\Omega^{\alpha_1\beta_1;\alpha_2\beta_2}
\frac{\partial\psi_a}{\partial z^{\alpha_1\beta_1}}
\frac{\partial\psi_b}{\partial \bar{z}^{\alpha_2\beta_2}}=
\sum_{c}f_{ab}^c\psi_c,\label{poiscoad}
\ee
where $\psi_{a,b,c}$ belong to a given irrep ${\cal H}_{\vec{v}}(G)$.
The structure constants for (\ref{poiscoad}) can be obtained through $f_{ab}^{c}=
\scprod{\psi_c}{\{\psi_a,\psi_b\}}$ when the set $\{{\psi}_a\}$ is
chosen to be orthonormal.
To each function $\psi\in C^\infty(F)$, one
can assign its {\it Hamiltonian vector field} ${H}_\psi\equiv\{\psi,\cdot\}$, which is
obviously divergence-free and preserves the natural volume form $\Omega^{N(N-1)/2}$. Thus,
the space Ham$(F)$ of Hamiltonian vector fields is, in general, a
subalgebra of the algebra sdiff$(F)$ of symplectic (volume-preserving)
diffeomorphisms of $F$, as stated in (\ref{classlim}). They only (essentially)
coincide in two dimensions, where $F_2=S^2$ and
$F_{1,1}=S^{1,1}$ (the sphere and the
hyperboloid) \cite{Bergshoeff2}.

The author is aware that some of the previous physical ideas
are still rather conjectural
and not quite developed or founded.
Nevertheless, the mathematical structure, as such,
is somewhat promising and deserves further study. A more explicit
interconnection between tensor operator, group, Poisson and
symplectic diffeomorphism algebras,
and non-commutative geometry on $U(N_+,N_-)$ is in progress \cite{progress}.

\section{Conclusions and outlook}

The long sought-for unification of all interactions and exact solvability of
(quantum) field theory and statistics parallels the quest for new symmetry
principles. Symmetry is an essential resource when facing
those two fundamental problems, either as a gauge guide principle or
as a valuable classification tool. The representation theory of infinite-dimensional
groups and algebras has not progressed very far, except for some important achievements
in one and two dimensions (mainly Virasoro and Kac-Moody symmetries), and
necessary breakthroughs in the subject remain to be carried out. This article
intends to fill just part of this gap by providing tractable higher-dimensional analogies
of the infinite two-dimensional conformal symmetry and its generalizations
in the context of higher-conformal-spin fields on anti-de Sitter spaces.
We have discussed the potential role of these new symmetry algebras
in the understanding of important aspects of (the still unknown) quantum gravity
like: radiation phenomena and quantum (noncommutative) structure of the space
at Planck scale. The next step will consist in an exhaustive study of the representations
of these tensor operator (and group) algebras and the construction of
invariant geometric action functionals. Other facets in which these symmetries
could be of use are: integrable (classical and quantum) nonlinear field
models and phase transitions in higher dimensions.

\appendix

\section{Appendix: Higher-$U(2,2)$-spin fields on AdS$_5$\label{intgeom}}

In order to put coordinates on AdS$_5$, the ideal choice is the
Bruhat cell decomposition \cite{Helgason} of $SU(2,2)$ for the
flag manifold $F_{2,2}=SU(2,2)/U(1)^3$ (e.g. the maximal coadjoint
orbit of $SU(2,2)$). According to this Bruhat decomposition, the
flag manifold $F_{2,2}$ can be covered with several coordinate
patches. We shall restrict ourselves to the largest cell which
provides a complex coordinatization $\{z_{\alpha\beta},
\alpha>\beta=1,2,3\}$ of nearly all $F_{2,2}$, missing only
lower-dimensional subspaces. A triangulation process
\be
\bordermatrix{&\u1 &\u2 &\u3 &\u4 \cr &u_{11}&u_{12}&u_{13}
&u_{14} \cr &u_{21}&u_{22}&u_{23}&u_{24} \cr
&u_{31}&u_{32}&u_{33}&u_{34} \cr &u_{41}&u_{42}&u_{43}&u_{44} }
\longrightarrow \bordermatrix{&\z1 &\z2 &\z3 &\z4 \cr & 1 & 0&0 &0
\cr & z_{21} & 1&0&0&\cr& z_{31}&z_{32}&1&0&\cr&
z_{41}&z_{42}&z_{43}&1}\label{triang} \ee gives us the coordinates
\bea &
&z_{21}=\frac{u_{21}}{u_{11}},\,z_{31}=\frac{u_{31}}{u_{11}},\,
z_{41}=\frac{u_{41}}{u_{11}},\nn\\
&&z_{32}=\frac{u_{11}u_{32}-u_{12}u_{31}}{u_{11}u_{22}-u_{12}u_{21}},\,
z_{42}=\frac{u_{11}u_{42}-u_{12}u_{41}}{u_{11}u_{22}-u_{12}u_{21}},
\label{su22coord}\\
&&z_{43}=\frac{u_{13}(u_{21}u_{42}-u_{22}u_{41})-
u_{23}(u_{11}u_{42}-u_{12}u_{41})
+u_{43}(u_{11}u_{22}-u_{12}u_{21})}{u_{13}(u_{21}u_{32}-
u_{22}u_{31})-u_{23}(u_{11}u_{32}-u_{12}u_{31})
+u_{33}(u_{11}u_{22}-u_{12}u_{21})},\nn \eea on the 6-dimensional
complex (non-compact) manifold $F_{2,2}$. These coordinates are
defined where the denominators are non-zero. Other patches
basically correspond to different locations of 0's and 1's in the
``triangulation'' process (\ref{triang}). The Cartan subgroup
$U(1)^3$ corresponds to diagonal matrices $h={\rm
diag}(h_1,h_2/h_1,h_3/h_2,1/h_3)$ with coordinates \bea
&&h_1=\left(\frac{u_{11}}{\bar{u}_{11}}\right)^{1/2},\,
h_2=\left(\frac{u_{11}u_{22}-u_{12}u_{21}}{\bar{u}_{11}\bar{u}_{22}-
\bar{u}_{12}\bar{u}_{21}}\right)^{1/2},\\
&&h_3=\left(\frac{u_{13}(u_{21}u_{32}-
u_{22}u_{31})-u_{23}(u_{11}u_{32}-u_{12}u_{31})
+u_{33}(u_{11}u_{22}-u_{12}u_{21})}{\bar{u}_{13}(\bar{u}_{21}\bar{u}_{32}-
\bar{u}_{22}\bar{u}_{31})-\bar{u}_{23}(\bar{u}_{11}\bar{u}_{32}-
\bar{u}_{12}\bar{u}_{31}) +\bar{u}_{33}(\bar{u}_{11}\bar{u}_{22}
-\bar{u}_{12}\bar{u}_{21})}\right)^{1/2}.\nn \eea Let us regard
any pseudo-unitary matrix $v\in SU(2,2)$ as a juxtaposition of
four column vectors $v=(\w1,\w2,\w3,\w4)$. The expression of $v$
in minimal coordinates $z_{\alpha\beta}$ can be obtained (up to
torus elements $h$) by means of an ``adapted'' Gramm-Schmidt
orthonormalization process of the set $\{\z\alpha\}$ in
(\ref{triang}) as follows:
\be
\vec{v}_\alpha'=\left(\z\alpha-
\frac{(\z\alpha,\vec{v}_{\alpha-1})}{(\vec{v}_{\alpha-1},\vec{v}_{\alpha-1})}
\vec{v}_{\alpha-1}-\dots -\frac{(\z\alpha,\w1)}{(\w1,\w1)}\w1\right),\;\;
\vec{v}_\alpha=\frac{\vec{v}_\alpha'}{
\left(\eta^{\alpha\alpha}(\vec{v}_\alpha',\vec{v}_\alpha')\right)^{1/2}},\label{g-s}
\ee
where $(\u\alpha,\vec{v}_\beta)\equiv
\bar{u}_{\alpha\mu}\eta^{\mu\nu}v_{\beta\nu}$
denotes a scalar product with indefinite metric $\eta={\rm diag}(1,1,-1,-1)$.
The explicit expression of (\ref{g-s}) proves to be:
\bea
\w1&=&\frac{1}{\Delta_1}\left(\ba{c} 1 \\ z_{21} \\ z_{31} \\ z_{41}\ea\right),\,\,
\w2=\frac{1}{\Delta_1\Delta_2}\left(\ba{c}
-{\bar{z}_{21}}+z_{32} {\bar{z}_{31}}+z_{42} {\bar{z}_{41}} \\
1+z_{32} z_{21} {\bar{z}_{31}}-z_{31} {\bar{z}_{31}}+z_{42}
z_{21} {\bar{z}_{41}}-z_{41} {\bar{z}_{41}} \\
z_{32}+z_{32} z_{21} {\bar{z}_{21}}-
{\bar{z}_{21}} z_{31}+z_{42} z_{31} {\bar{z}_{41}}-
z_{32} z_{41} {\bar{z}_{41}} \\ z_{42}+z_{42}
z_{21} {\bar{z}_{21}}-z_{42} z_{31} {\bar{z}_{31}}-
{\bar{z}_{21}} z_{41}+z_{32} {\bar{z}_{31}} z_{41}\ea  \right)
\nn\\
\w3&=&\frac{1}{\Delta_2\Delta_3}\left(\ba{c} \left[
-{\bar{z}_{32}} {\bar{z}_{21}}-
{\bar{z}_{42}} z_{43} {\bar{z}_{21}}+{\bar{z}_{31}}-z_{42} {\bar{z}_{42}}
{\bar{z}_{31}}\right.\\
\left.+z_{32} {\bar{z}_{42}} z_{43} {\bar{z}_{31}}+{\bar{z}_{32}}
z_{42} {\bar{z}_{41}}+z_{43} {\bar{z}_{41}}-z_{32}
{\bar{z}_{32}} z_{43} {\bar{z}_{41}}\right] \\
\left[{\bar{z}_{32}}+{\bar{z}_{42}} z_{43}-z_{42} {\bar{z}_{42}}
z_{21} {\bar{z}_{31}}+z_{32} {\bar{z}_{42}} z_{43}z_{21}
{\bar{z}_{31}}-{\bar{z}_{42}} z_{43} z_{31} {\bar{z}_{31}}+
{\bar{z}_{42}} {\bar{z}_{31}} z_{41}\right.\\
\left.+{\bar{z}_{32}}z_{42} z_{21} {\bar{z}_{41}}-
z_{32} {\bar{z}_{32}} z_{43} z_{21} {\bar{z}_{41}}+
{\bar{z}_{32}} z_{43} z_{31} {\bar{z}_{41}}-
{\bar{z}_{32}} z_{41} {\bar{z}_{41}}\right] \\
\left[1-z_{42} {\bar{z}_{42}}+z_{32}{\bar{z}_{42}} z_{43}-
z_{42} {\bar{z}_{42}} z_{21} {\bar{z}_{21}}+
z_{32} {\bar{z}_{42}} z_{43} z_{21} {\bar{z}_{21}}-
{\bar{z}_{42}}z_{43} {\bar{z}_{21}} z_{31}\right.\\
\left.+  {\bar{z}_{42}} {\bar{z}_{21}} z_{41}+z_{42} z_{21}
 {\bar{z}_{41}}-z_{32} z_{43} z_{21} {\bar{z}_{41}}+z_{43} z_{31}
{\bar{z}_{41}}-z_{41} {\bar{z}_{41}} \right]
\\ \left[{\bar{z}_{32}} z_{42}+z_{43}-z_{32} {\bar{z}_{32}} z_{43}+
{\bar{z}_{32}} z_{42} z_{21} {\bar{z}_{21}}-z_{32} {\bar{z}_{32}} z_{43} z_{21}
 {\bar{z}_{21}}+{\bar{z}_{32}} z_{43} {\bar{z}_{21}} z_{31}\right.\\
\left.-z_{42} z_{21}{\bar{z}_{31}}+z_{32} z_{43} z_{21} {\bar{z}_{31}}
-z_{43} z_{31} {\bar{z}_{31}}-{\bar{z}_{32}} {\bar{z}_{21}}
z_{41}+{\bar{z}_{31}}z_{41}\right]
\ea \right)
\nn\\
\w4&=&\frac{1}{\Delta_3}\left(\ba{c}-{\bar{z}_{42}} {\bar{z}_{21}}+{\bar{z}_{32}} {\bar{z}_{43}}
{\bar{z}_{21}}-{\bar{z}_{43}}
{\bar{z}_{31}}+{\bar{z}_{41}}\\ {\bar{z}_{42}}-{\bar{z}_{32}}
{\bar{z}_{43}}\\ -{\bar{z}_{43}}\\ 1\ea\right)
\eea
where
\begin{eqnarray}
\Delta_1(z,\bar{z})&=&\sqrt{1+|z_{21}|^2-|z_{31}|^2 -|z_{41}|^2}
\label{lengths}\\
\Delta_2(z,\bar{z})&=&\sqrt{1+|z_{32}z_{41}-z_{42}z_{31}|^2-|z_{32}|^2-
|z_{42}|^2 -|z_{32}z_{21}-z_{31}|^2-|z_{42}z_{21}-z_{41}|^2}
\nn\\
\Delta_3(z,\bar{z})&=&\sqrt{1+|z_{43}|^2-|z_{42}-z_{43}z_{32}|^2-|z_{41}+
z_{43}z_{32}z_{21}-z_{42}z_{21}-z_{43}z_{31}|^2}\nn
\end{eqnarray}
are three characteristic lengths that will play a central role in what
follows.

Using a relativistic notation, we may say that the vectors $\w1$
and $\w2$ are ``space-like'' (that is,
$(\vec{v}_{1,2},\vec{v}_{1,2})=1$) whereas $\w3$ and $\w4$ are
``time-like'' (i.e, $(\vec{v}_{3,4},\vec{v}_{3,4})=-1$); this
ensures that $v\eta v^\dag=\eta$. Any $u\in SU(2,2)$ matrix in the
present patch (which contains the identity element
$z=0=\bar{z},h=1$) can be written in minimal coordinates
$z_{\alpha\beta}, \bar{z}_{\alpha\beta},
h_\beta,\,\alpha>\beta=1,2,3$ as the product $u=vh$ of an element
$v$ on the base $F_{2,2}$ times an element $h$ on the fiber
$U(1)^3$.

Once we have the expression of a general $SU(2,2)$ group
element $g=u$ in terms of the minimal coordinates
$u=u(z_{\alpha\beta},\bar{z}_{\alpha\beta},h_\beta)$,
we can easily write the group
law $u''=u'\bullet u$ and compute the left- and right-invariant vector fields
(\ref{izquierdo},\ref{derecho}),
as for the $SU(1,1)$ case (\ref{law1},\ref{lrivf}).
We shall not give the (rather cumbersome) explicit
expression of the, let us say,
right-invariant vector fields, but the relation between them and the
step operators (\ref{step}) with commutation relations
(\ref{pun}). This correspondence is:
\be
X^{R}_{z_{\alpha\beta}}\to{X_{\alpha}}^\beta,\;\;
X^{R}_{\bar{z}_{\alpha\beta}}\to-{X^\beta}_{\alpha},\;\;
X^{R}_{h_{\beta}}\to{X_{\beta}}^\beta-{X_{\beta+1}}^{\beta+1},\;\;\;
\alpha>\beta=1,2,3.
\ee
The Casimir operators for $SU(2,2)$ are easily written in the basis of
step operators as follows:
\be
C_2={X_{\alpha}}^\beta{X_{\beta}}^\alpha,\;\;
C_3={X_{\alpha}}^\beta{X_{\beta}}^\gamma{X_{\gamma}}^\alpha,\;\;
C_4={X_{\alpha}}^\beta{X_{\beta}}^\gamma{X_{\gamma}}^\sigma
{X_{\sigma}}^\alpha. \label{casisun}
\ee
where the trace $C_1={X_{\alpha}}^\alpha$ is zero for special groups.
Associated with the previous three Casimir operators,
there are three possible choices of time-generators $X^L_{t_\beta}\equiv
\frac{\omega_\beta}{2}X^L_{h_\beta}$
to construct the Hilbert space
${\cal H}_S(SU(2,2))$ of holomorphic (i.e. $X^L_{z}\psi=0$)
$SU(2,2)$-spin $S$ wave functions $\psi^{(S)}$ on AdS$_5$. The general solution to the equations
of motion proves to be:
\be
\left.\ba{l} X^L_{t_\beta}\psi=-\omega_\beta S_\beta\psi \\
X^L_{z_{\alpha\beta}}\psi=0\ea\right\}\Rightarrow
\psi^{(S)}(u)=W^{(S)}(h,z,\bar{z})\varphi(\bar{z}), \label{antiholf}
\ee
where
\be W^{(S)}(h,z,\bar{z})\equiv \prod_{\beta=1}^3
\left(h_\beta \Delta_\beta(z,\bar{z})\right)^{-2S_\beta},\;\;\;
\varphi(\bar{z})\equiv \sum_{m} a^{(S)}_{m}\prod_{\alpha>\beta}
(\bar{z}_{\alpha\beta})^{m_{\beta\alpha}}\label{vacsu22}
\ee
represents the vacuum wave function $W^{(S)}$, which is written
in terms of the toral (angular) coordinates $h_\beta$ and the
typical lengths $\Delta_\beta$ in (\ref{lengths}), and $\varphi$ is
an analytic power series,
with complex coefficients $a^{(S)}_{m}$,  on its arguments
$\bar{z}_{\alpha\beta}$, respectively. The index $m$ denotes an integral
upper-triangular $4\times 4$ matrix (see Eq. (\ref{uppertmatrix})) and
$S=(S_1,S_2,S_3)$ is the $SU(2,2)$-spin three-dimensional vector which
lies on a half-integer lattice.
Let us discuss the range of variation of the $S_\beta$ and $m_{\alpha\beta}$
indices.

The sign of the $SU(2,2)$-spin indices $S_\beta$
depends on the (non-)compact character of the
corresponding simple roots
(the ones whose generators $X_{\alpha\beta}$ are labeled by
$\alpha\beta=12,23,34$). In this notation,
the roots $\alpha\beta=12,34$ are of
compact type, which implies $S_1,S_3\in\Z^+/2$, and the root
$\alpha\beta=23$ is of non-compact type, which leads to $S_2\in\Z^-/2$.
This also guarantees the finiteness of the scalar product
(\ref{scalarprod}) with left-invariant integration measure
\be
d^Lu=\frac{i}{(2\pi)^6}\frac{1}{\prod_{\beta=1}^3(\Delta_\beta)^4}
\bigwedge_{\beta=1}^{3}h_\beta^{-1}dh_\beta
\bigwedge_{\alpha>\beta}d\Re(z_{\alpha\beta})\wedge
d\Im(z_{\alpha\beta}), \ee and the unitarity of the representation
$\rho(u')\psi^{(S)}(u)= \psi^{(S)}(u'^{-1}\bullet u)$ of
$SU(2,2)$. We can still keep track of the extra $U(1)$ quantum
number $S_4$ that differentiates $U(2,2)\simeq (SU(2,2)\times
U(1))/\Z_4$ from $SU(2,2)$ representations. The $U(2,2)$ wave
functions $\tilde{\psi}^{(I)}$ depend on an extra $U(1)$-factor
$(h_4)^{-2S_4}, h_4\in U(1)$ in the vacuum wave function $W^{(S)}$
in (\ref{vacsu22}), where the relation between the $U(2,2)$-spin
labels $I=(I_1,I_2,I_3,I_4)$ of Sec. (\ref{v3p1}) and the
$SU(2,2)\times U(1)$-spin labels $S=(S_1,S_2,S_3,S_4)$ is
$S_\beta=I_\beta-I_{\beta+1}, \beta=1,2,3$ and
$S_4=I_1+I_2+I_3+I_4$ (the Casimir $C_1$ ---trace--- eigenvalue).
For completeness, we shall give the action of the Cartan
subalgebra (Hamiltonian operators) on basic wave functions
$\psi^{(S)}_m\equiv
W^{(S)}\prod_{\alpha>\beta}(\bar{z}_{\alpha\beta})^{m_{\beta\alpha}}$
of ${\cal H}_S(SU(2,2))$. It can be easily calculated from the
explicit expression of the right-invariant differential operators
$X^R_{h_\beta}$ and is proven to be:
\be
X^R_{h_\beta}\psi^{(S)}_m=(2S_\beta+m_\beta-m_{\beta+1})\psi^{(S)}_m,
\label{cartanact}
\ee
where $m_\beta$ is defined after Eq. (\ref{uppertmatrix}); note that
the eigenvalue $(2S_\beta+m_\beta-m_{\beta+1})$ of $X^R_{h_\beta}$
can also be written as
$2(\Gamma^+_\beta-\Gamma^+_{\beta+1})$, where
$\Gamma^\pm_\beta\equiv I_\beta \pm m_\beta/2$ is a characteristic
quantity that plays a central role (for example, the structure
constants of the algebra (\ref{inf}) and its higher order
quantum corrections (\ref{infq}) could also be written as powers of
$\Gamma$'s). It is also worth noticing that, from (\ref{cartanact}), the
quantities  $E_\beta^{(0)}\equiv \hbar\omega_\beta S_\beta$
represent the vacuum expectation values (zero-point energies)
of the Hamiltonian operators $X^R_{t_\beta}$ defined
above. The action of the remainder operators $X^R_{z,\bar{z}}$ on basic
functions can also be calculated in the same way, although we shall not
give the explicit expression here.

In order to find the domain of the matrix indices $m_{\alpha\beta}$ for
each set of spin labels $\{S_\beta\}$, an ``orbit-through-the-vacuum''
procedure could be used, which consists in an iterative application of
raising (creation) operators $X^R_{z_{\alpha\beta}}$  on the
vacuum $\psi^{(S)}_0$. Another possibility is to look at the
propagator $\Delta^{(S)}(u,u')$, whose particular form can be
inferred from the expression (\ref{propaholo}) for the
AdS$_2$ case; indeed, it can be written in terms of the lengths $\Delta_\beta$
as follows:
\be
\Delta^{(S)}(u,u')=\sum_m\frac{1}{N^{(S)}_m}\psi^{(S)}_m(u)
\bar{\psi}^{(S)}_m(u')=\kappa_S\prod_{\beta=1}^3
\frac{\Delta_\beta(z',\bar{z})^{4S_\beta}}{(h_\beta
\Delta_\beta(z,\bar{z}))^{2S_\beta}(\bar{h}'_\beta
\Delta_\beta(z',\bar{z}'))^{2S_\beta}},\label{propaholo2}
\ee
where $N^{(S)}_m$ are the normalization (squared) constants of the basic
functions  $\psi^{(S)}_m$ and $\kappa_S$ is a global constant depending on
$S$. The expansion of the factor $\Delta_\beta(z',\bar{z})^{4S_\beta}$ in
(\ref{propaholo2}) in powers of $z'_{\alpha\beta}\bar{z}_{\alpha\beta}$
tells us which $m_{\beta\alpha}$ are present in the summation
(\ref{propaholo2}) ---and (\ref{vacsu22})--- for a given $SU(2,2)$-spin
$S$. Taking into account that $S_3\leq 0$, it is clear that some of the
$m_{\alpha\beta}$ indices have no upper-bound, as  corresponds
in general to unitary irreducible representations of non-compact groups.

Before concluding this Appendix, let us briefly comment on K\"ahler
geometric structures on the flag manifold $F_{2,2}$ (it also applies
to general $F_{{}_{N_+,N_-}}$). Since the flag manifold $F_{2,2}$ is a
K\"ahler manifold \cite{Nakahara}, it possesses complex local
coordinates $z_{\alpha\beta}$ (\ref{su22coord}),
an Hermitian Riemannian metric $g$ and a
corresponding  closed two-form (K\"ahler form) $\Omega$
\be
ds^2=g^{\alpha\beta,\mu\nu} dz_{\alpha\beta}d\bar{z}_{\mu\nu},\;\;
\Omega=i g^{\alpha\beta,\mu\nu}dz_{\alpha\beta}\wedge d\bar{z}_{\mu\nu},\label{kahlerform}
\ee
which can be obtained from the K\"ahler potential
\be
K^{(S)}(z,\bar{z})=\sum_{\beta=1}^3 2S_\beta\ln
\Delta_\beta(z,\bar{z}) \label{kahlerpot} \ee through the formula
$g^{\alpha\beta,\mu\nu}=\frac{\partial}{\partial z_{\alpha\beta}}
\frac{\partial}{\partial \bar{z}_{\mu\nu}}K^{(S)}$, for a given
coadjoint orbit of $SU(2,2)$ characterized by the indices
$S_\beta$. Keep in mind that, in complex differential calculus,
$\partial^2=0=\bar{\partial}^2$ and
$\partial\bar{\partial}+\bar{\partial}\partial=0$. Note also that,
the K\"ahler potential $K^{(S)}(z,\bar{z})$ essentially
corresponds to the natural logarithm of the vacuum
$W^{(S)}(h,z,\bar{z})$ in (\ref{vacsu22}), up to toral coordinates
$h$.

\section*{Acknowledgments}

I wish to acknowledge the ``Consejer\'\i a de Educaci\'on y Ciencia,
Junta de Andaluc\'\i a" for a grant-in-aid.
I also would like to thank S. Howes for bringing to my attention the Bruhat cell
decomposition and for his interest, comments and valuable help;
I also extend my gratitude to J.L. Jaramillo. Work partially supported by
the DGICYT under project PB98-0520.

\end{document}